\documentclass[aps,prb,twocolumn,showpacs]{revtex4}

\usepackage[T1]{fontenc}
\usepackage{amssymb}
\usepackage{amsbsy}
\usepackage{amsmath}           
\usepackage{epsfig}

\begin{document}


\def\beq{\begin{equation}}
\def\eeq{\end{equation}}
\def\bleq{\begin{eqnarray}}
\def\eleq{\end{eqnarray}} 
\newcommand{\Tr}{{\rm Tr}} 
\newcommand{\mean}[1]{\langle #1 \rangle}

\def\down{\downarrow}
\def\eps{\epsilon}
\def\gam{\gamma} 
\def\Ome{\Omega}
\def\bfOme{\boldsymbol{\Omega}} 
\def\sig{\sigma}
\def\bfsig{\boldsymbol{\sigma}} 
\def\The{\Theta} 
\def\up{\uparrow}

\def\xik{\xi_{\bf k}} 
\def\Ek{E_{\bf k}}

\def\half{\frac{1}{2}}
\def\quarter{\frac{1}{4}}

\def\a{{\bf a}}
\def\b{{\bf b}}
\def\k{{\bf k}}
\def\n{{\bf n}} 
\def\p{{\bf p}} 
\def\q{{\bf q}}
\def\r{{\bf r}}
\def\t{{\bf t}}
\def\v{{\bf v}}
\def\A{{\bf A}}
\def\H{{\bf H}}  
\def\J{{\bf J}}
\def\K{{\bf K}} 
\def\L{{\bf L}}
\def\M{{\bf M}}  
\def\P{{\bf P}} 
\def\Q{{\bf Q}} 
\def\R{{\bf R}}
\def\S{{\bf S}}
\def\bfnabla{\boldsymbol{\nabla}}
\def\bfsigma{\boldsymbol{\sigma}} 

\def\para{\parallel}
\def\kpara{{k_\parallel}}

\def\w{\omega}
\def\wn{\omega_n}
\def\wnu{\omega_\nu}
\def\dtau{{\partial_\tau}} 

\def\intk{\int_{\bf k}} 
\def\intr{\int d^3r\,} 
\def\intrp{\int d^3r'\,}
\def\intt{\int_0^\beta d\tau}
\def\inttp{\int_0^\beta d\tau'}
\def\intx{\int d^4x\,} 
\def\inttr{\int_0^\beta d\tau \int d^3r\,}
\def\intinf{\int_{-\infty}^\infty}

\def\calD{{\cal D}}
\def\calG{{\cal G}}
\def\calO{{\cal O}}

\def\Phid{\Phi^{(2)}} 
\def\Phipp{\Phi^{(2)}_{\rm pp}}
\def\Phiph{\Phi^{(2)}_{\rm ph}}
\def\Phit{\Phi^{(2)}_{\rm t}} 
\def\Phis{\Phi^{(2)}_{\rm s}} 
\def\Phich{\Phi^{(2)}_{\rm ch}} 
\def\Phisp{\Phi^{(2)}_{\rm sp}} 
\def\PhiCt{\Phi^{\rm C}_{\rm t}}
\def\PhiCs{\Phi^{\rm C}_{\rm s}}
\def\PhiPch{\Phi^{\rm P}_{\rm ch}}
\def\PhiPsp{\Phi^{\rm P}_{\rm sp}}
\def\PhiLch{\Phi^{\rm L}_{\rm ch}}
\def\PhiLsp{\Phi^{\rm L}_{\rm sp}}
\def\gammaCt{\gamma^{\rm C}_{\rm t}}
\def\gammaCs{\gamma^{\rm C}_{\rm s}}
\def\gammaPch{\gamma^{\rm P}_{\rm ch}}
\def\gammaPsp{\gamma^{\rm P}_{\rm sp}}
\def\gammaLch{\gamma^{\rm L}_{\rm ch}}
\def\gammaLsp{\gamma^{\rm L}_{\rm sp}}


\title{Renormalization group approach to interacting fermion systems in the
  two-particle-irreducible formalism} 
\author{N. Dupuis}
\affiliation{Department of Mathematics, Imperial College, 
180 Queen's Gate, London SW7 2AZ, UK}
\affiliation{Laboratoire de Physique des Solides, CNRS UMR 8502,
  Universit\'e Paris-Sud, 91405 Orsay, France}
 
\date{June 21, 2005}
\begin{abstract}  
We describe a new formulation of the functional renormalization group (RG) for
interacting fermions within a Wilsonian momentum-shell approach. We show that
the Luttinger-Ward functional is invariant under the RG transformation, and
derive the infinite hierarchy of flow equations satisfied by  
the two-particle-irreducible (2PI) vertices. In the one-loop 
approximation, this hierarchy reduces to two equations that determine 
the self-energy and the 2PI two-particle vertex $\Phi^{(2)}$. 
Susceptibilities are calculated from the Bethe-Salpeter equation that relates
them to $\Phi^{(2)}$. While the one-loop approximation breaks down at low
energy in one-dimensional
systems (for reasons that we discuss), it reproduces the exact results both in
the normal and ordered phases in single-channel (i.e. mean-field) theories, as
shown on the example of BCS theory. The possibility to continue the RG flow
into broken-symmetry phases is an essential feature of the 2PI RG scheme and
is due to the fact that the 2PI two-particle vertex, contrary to its 1PI
counterpart, is not singular at a phase transition. Moreover, the normal phase
RG equations can be directly used to
derive the Ginzburg-Landau expansion of the thermodynamic potential near a
phase transition. We discuss the implementation of the 2PI RG scheme to
interacting fermion systems beyond the examples (one-dimensional systems and
BCS superconductors) considered in this paper. 
\end{abstract}

\maketitle

\section{Introduction}

The two-particle-irreducible (2PI) formalism
\cite{Luttinger60,Baym61,Baym62,Dedominicis64,note3} was
first introduced in condensed-matter physics as a means to systematically set
up self-consistent approximations that satisfy conservation laws. It can be
cast in a variational framework where the thermodynamic potential $\Gamma$ is
expressed as a functional of the single-particle Green function.\cite{note4} 
$\Gamma$ is essentially determined by the Luttinger-Ward (LW) functional
$\Phi$,\cite{Luttinger60} given by the sum of the 2PI Feynman
diagrams. $\Phi$ is also the generating functional 
of the self-energy and higher-order 2PI vertices. The so-called
$\Phi$-derivable approximations are based on truncations of the diagrammatic
expansion of $\Phi$ that retain only a finite number or a sub-series of
diagrams. They are thermodynamically consistent and satisfy conservation
laws.\cite{Baym62,note5} While most applications of 
the 2PI formalism to interacting fermion systems have been limited to the
Hartree-Fock level, recent 
developments, motivated by the physics of high-temperature superconductors,
have incorporated exchange of spin fluctuations within the Hubbard
model.\cite{Bickers89a,Bickers89b,Vilk97} Following general ideas put forward
by Wetterich,\cite{Wetterich02} the aim of this paper is to discuss a 2PI
formulation of the renormalization group (RG) approach to interacting fermion
systems. 

The RG has proven a powerful approach for studying low-dimensional
fermion systems, providing a systematic and unbiased method to study
competing instabilities in the weak-coupling limit
(Refs.~\onlinecite{Bourbonnais95}-\onlinecite{Freire05}).  
One of its main successes has been to explain how unconventional
superconductivity can occur at low temperature in systems like
organic conductors
\cite{Bourbonnais95,Duprat01,Bourbonnais04,Nickel05a,Nickel05b} 
or high-temperature superconductors 
\cite{Zanchi98,Zanchi00,Halboth00,Honerkamp01,Salmhofer01} where the dominant  
electron-electron interactions are expected to be repulsive and
favor antiferromagnetism. Although the RG can be implemented in different
ways, most approaches rely on the so-called one-particle-irreducible (1PI)
RG scheme or variants thereof.\cite{Honerkamp01,Salmhofer01,note6} The 1PI RG
scheme is based on an exact RG equation for the generating functional of 1PI
vertices. The existence of a Fermi surface implies that the interaction
amplitudes 
strongly  depend on the momenta of the interacting fermions, which leads to
functional RG equations for the 1PI vertices. For this reason, most RG
calculations in fermion systems have been limited to one-loop
order and are thus restricted to the weak-coupling limit. Another severe
limitation of the method comes from the difficulty to 
access broken-symmetry phases. Long-range order is signaled by a diverging
flow of certain 1PI vertices and susceptibilities at a critical energy
or temperature scale, below which the RG equations cannot be continued. It has
been proposed to circumvent this difficulty by introducing an infinitesimally
small symmetry-breaking component in the initial condition of the RG
equations,\cite{Salmhofer05} or combining the RG technique with a mean-field
approximation at low energy.\cite{Fuseya05a,Metzner05}
Alternatively, one can introduce a partial
bosonization of the action yielding a description in terms of both fermionic
and collective Hubbard-Stratonovich fields.
\cite{Wetterich02,Baier04,Baier05,Schutz05} It will be interesting to further
explore the applicability of these new RG schemes in various models. 

The main purpose of this paper is to show that broken-symmetry phases can be
studied with a RG scheme where the basic quantities are the 2PI
vertices rather than their 1PI counterparts. To understand this issue in
simple terms, let us consider the 1PI two-particle vertex in the spin singlet
particle-particle channel obtained by summing the bubble diagrams (random-phase
approximation (RPA)),
\beq
\gamma = \frac{\gamma_0}{1+ l \gamma_0} ,
\label{ex1}
\eeq
where $l\sim \ln(\Lambda_0/\Lambda)$ comes from the non-interacting
particle-particle propagator. $\Lambda_0$ is a high-energy cutoff
(e.g. the bandwidth), $\Lambda$ an infrared cutoff which can be identified
with the temperature, and $\gamma_0$ the bare (dimensionless) interaction. 
In the RG framework, $\gamma$ can be interpreted either as the 1PI vertex of a
theory with infrared cutoff $\Lambda$ or as the effective interaction of the
Wilsonian action with ultraviolet cutoff $\Lambda$.\cite{Morris94}  
Eq.~(\ref{ex1}) becomes a differential equation (see
e.g. Ref.~\onlinecite{Shankar94}),  
\beq
\frac{d\gamma}{dl} = - \gamma^2 .
\label{ex2}
\eeq
For an attractive interaction $\gamma_0<0$, a divergence occurs at the energy
(or temperature) scale $\Lambda_c=\Lambda_0\exp(1/\gamma_0)$, signaling the 
formation of Cooper pairs and the appearance of long-range superconducting
order. In the BCS (mean-field) 
theory, this divergence is cured below $\Lambda_c$ by the presence of a
finite gap in the fermion excitation spectrum. However, in the 1PI RG scheme, 
it prevents the flow to be straightforwardly\cite{Salmhofer05} continued into
the broken-symmetry phase.  

Let us now reconsider Eq.~(\ref{ex1}) from a different point of view. This
equation can be seen as a Bethe-Salpeter
equation in the particle-particle channel with the bare interaction $\gamma_0$
as the 2PI vertex $\Phid$ (the reason for this notation will become clear in
Sec.~\ref{sec:formalism}). Only the non-interacting particle-particle
propagator $\sim l\sim \ln(\Lambda/\Lambda_0)$ is scale dependent in
Eq.~(\ref{ex1}), while the 2PI vertex $\Phid$ is invariant under
the RG transformation,
\beq
\frac{d\Phid}{dl} = 0 .
\eeq 
Within the BCS theory, the 2PI vertex $\Phid$ is therefore not sensitive to the
transition into the superconducting phase. The appearance of long-range order
is expected to induce an anomalous (i.e. symmetry violating) self-energy below
$\Lambda_c$.  
In more complicated cases, where several types of fluctuations may compete
together, we cannot exclude the appearance of singularities in the 2PI
vertices. For instance, in a conductor close to an antiferromagnetic
instability, singularities in the particle-particle channel (i.e. in the
particle-particle component of $\Phid$) may be induced by nearly divergent
spin fluctuations. We shall discuss this point in the concluding section and
explain how these singularities can be controlled by a proper parameterization
of the vertex. All these considerations suggest to use a RG scheme where the
basic objects are the 2PI vertices.  

The outline of the paper is as follows. In Sec.~\ref{subsec:LW}, we briefly
recapitulate the 2PI formalism. By means of a Legendre transformation, 
we express the thermodynamic potential (grand potential) as a functional
$\Gamma[G]$ of the 
Green function; we then  define the LW functional $\Phi[G]$ and the 2PI
vertices $\Phi^{(n)}$. In Sec.~\ref{subsec:rgeq_gen}, we describe the RG
procedure. We derive the differential equation satisfied by the thermodynamic
potential and show that the LW functional is invariant under the RG
transformation. We then deduce the infinite hierarchy of flow equations
satisfied by the 2PI vertices. A one-loop approximation is then introduced by
truncating this hierarchy and approximating the 2PI three-particle vertex in
terms of the 2PI two-particle vertex $\Phid$ (Sec.~\ref{subsec:rgeq_1L}).
We discuss the connection between the one-loop equations and their counterparts
in the 1PI RG scheme. In Secs.~\ref{subsec:rgeq_N} and
\ref{subsec:RFN}, we give explicit expressions of the one-loop flow
equations in the normal phase and discuss the
calculation of response functions. Some of the 
general results of Secs.~\ref{subsec:LW}-\ref{subsec:RFN} have been
previously obtained by Wetterich,\cite{Wetterich02} sometimes in a slightly
different formulation, with a few important differences that we shall
mention. In Sec.~\ref{subsec:GL}, we show how the normal phase RG equations
can be used to derived the Ginzburg-Landau expansion of the thermodynamic
potential in the vicinity of a phase transition. This is achieved by
considering the Legendre transform $F[\Sigma]$ of the LW functional $\Phi[G]$,
which allows one to 
express the thermodynamic potential as a functional of the self-energy. The
latter is then split into a normal part and an anomalous (i.e. symmetry
violating) part which is to be determined by minimizing the thermodynamic
potential. The last two sections are
devoted to the application of the 2PI RG formalism to two different models. 
In Sec.~\ref{sec:BCS}, we consider a three-dimensional fermion system with an
attractive interaction. From the one-loop approximation restricted to the
particle-particle channel, we rederive the main results of the BCS theory (gap
equation, thermodynamic potential, and collective modes), thus showing the
ability of the 2PI RG scheme to access broken-symmetry phases. We find that
the 2PI two-particle vertex $\Phid$ is invariant under the RG transformation,
while the flow equation for the self-energy 
yields the BCS gap equation. In Sec.~\ref{sec:LL}, we study one-dimensional
(1D) systems within the g-ology framework. At one-loop order, the  
2PI scheme compares favorably to the RG scheme at high energy, but
deteriorates at lower energy and eventually breaks down. We identify the
reason for this failure and argue that the 2PI RG scheme can nevertheless be
applied to realistic quasi-1D systems like the organic conductors of the
Bechgaard salt family. The conclusion is devoted to a 
discussion of the implementation of the 2PI RG formalism beyond the
examples considered in Secs.~\ref{sec:BCS} and \ref{sec:LL}. 

RG approaches in the 2PI formalism have also been discussed in high-energy
physics in a field-theoretical framework (see, for instance,
Refs.~\onlinecite{VanHees02a,VanHees02b,VanHees02c,Blaizot04}).

\section{2PI RG formalism}
\label{sec:formalism}

\subsection{Luttinger-Ward functional and 2PI vertices}
\label{subsec:LW}

We consider the partition function of a spin-$\half$ fermion system in the
presence of external sources, 
\beq
Z[J] = \int \calD[\psi] \exp\Bigl\lbrace - S[\psi] + \half \psi^T J \psi
\Bigr\rbrace ,
\label{partition} 
\eeq 
where the action $S=S_0+S_{\rm int}$ is defined by  
\bleq
S_0[\psi] &=&  \half \sum_{\alpha,\beta}  \psi_\alpha C^{-1}_{\alpha\beta}
\psi_\beta , \nonumber \\ 
S_{\rm int}[\psi] &=& \frac{1}{4!} \sum_{\alpha_1,\alpha_2,\alpha_3,\alpha_4}
V_{\alpha_1\alpha_2\alpha_3\alpha_4} \psi_{\alpha_1} \psi_{\alpha_2} 
\psi_{\alpha_3} \psi_{\alpha_4} , 
\eleq  
with $C_{\alpha\beta}$ the free propagator and 
$V_{\alpha_1\alpha_2\alpha_3\alpha_4}$ the totally antisymmetrized
interaction vertex. $\half \psi^T J \psi =  \half \sum_{\alpha,\beta}
\psi_\alpha J_{\alpha\beta} \psi_\beta$ describes the
coupling to the external bosonic sources $J_{\alpha\beta}$. The
$\psi_\alpha$'s are Grassmann variables and the collective index $\alpha\equiv
(\r,\tau,\sig,c)$ labels the position, imaginary time and spin projection
along a given axis, as well as other possible internal degrees of freedom. 
$\sum_\alpha=\intt\int d^dr\sum_{\sig,c}$ where $\beta=1/T$ is the inverse
temperature and $d$ the space dimension. $c=\pm$ is a charge index such that 
\beq 
\psi_\alpha = \left\lbrace 
\begin{array}{ccc}
\psi_\sig(\r,\tau) & {\rm if} & c=- , \\  
\psi^*_\sig(\r,\tau) & {\rm if} & c=+ . 
\end{array}
\right . 
\eeq
Since the $\psi$'s anticommute, both $C_{\alpha\beta}$ and
$J_{\alpha\beta}$ can be chosen to be antisymmetric functions:
$C_{\alpha\beta}=-C_{\beta\alpha}$ and $J_{\alpha\beta}=-J_{\beta\alpha}$. In
the following, we denote by  
\beq
\gam = \lbrace \alpha,\beta \rbrace
\eeq
bosonic indices obtained from two fermionic indices $\alpha$ and $\beta$. 

The single-particle Green function is given by the functional derivative
of $W[J]=\ln Z[J]$,
\beq
G_\gam = \mean{\psi_\alpha\psi_\beta} = \frac{\delta W[J]}{\delta J_\gam} .
\label{G_def}
\eeq
Note that the definition of $G$ differs by a minus sign from the usual
definition in condensed-matter physics. The Legendre transform of $W[J]$ is
defined by  
\beq
\Gamma [G] = - W[J] - \half \Tr (JG) ,
\label{Gamma_def}
\eeq
where $J\equiv J[G]$ is obtained by inverting Eq.~(\ref{G_def}). To keep the
notations simple, we shall denote $J[G]$ by $J$ in the following. $\Tr$
denotes the trace with respect to the fermionic indices,
i.e. $\Tr(JG)=\sum_{\alpha,\beta} J_{\alpha\beta}G_{\beta\alpha}$. $\Gamma[G]$
satisfies the ``equation of motion''
\beq
\frac{\delta \Gamma[G]}{\delta G_\gam} = J_\gam ,
\label{state_eq}
\eeq
as can be easily verified by a direct calculation.\cite{note1}
It is customary to write $\Gamma[G]$ as
\beq
\Gamma[G] = \half \Tr \ln G - \half \Tr(GC^{-1}-1) + \Phi[G] , 
\label{phi_def}
\eeq  
where the LW functional $\Phi[G]$ is the sum of 2PI vacuum
fluctuation diagrams (or vacuum fluctuation skeleton diagrams), i.e. diagrams
that cannot be separated into two disconnected pieces by cutting two lines. 

By differentiating Eq.~(\ref{state_eq}) with respect to the source $J$ and
using Eq.~(\ref{G_def}),\cite{note1} 
we obtain  
\bleq
\bigl(\Gamma^{(2)} W^{(2)} \bigr)_{\gam_1\gam_2} &\equiv& 
\half \sum_{\gam_3} \Gamma^{(2)}_{\gam_1\gam_3}
W^{(2)}_{\gam_3\gam_2} \nonumber \\ 
&=& \delta_{\alpha_1,\alpha_2} \delta_{\beta_1,\beta_2} -
\delta_{\alpha_1,\beta_2} \delta_{\beta_1,\alpha_2} \nonumber \\ 
&\equiv& I_{\gam_1\gam_2}
\label{Gamma2_def}
\eleq
where
\bleq
\Gamma^{(n)}_{\gam_1\cdots\gam_n} &=& \frac{\delta^{(n)} \Gamma[G]}{\delta
  G_{\gam_1} \cdots \delta G_{\gam_n}} , \nonumber \\ 
W^{(n)}_{\gam_1\cdots\gam_n} &=& \frac{\delta^{(n)} W[J]}{\delta
  J_{\gam_1} \cdots \delta J_{\gam_n}} 
\label{Gamman_def}
\eleq 
are functionals of $G$. 
Eq.~(\ref{Gamma2_def}) defines a matrix multiplication with respect to the
bosonic indices with $I$ the unit matrix. Further relations between $\lbrace
W^{(n)}\rbrace$ and $\lbrace \Gamma^{(n)}\rbrace$ can be obtained by taking
higher-order derivatives of the equation of motion (\ref{state_eq}). 

The 2PI vertices are defined by  
\beq
\Phi^{(n)}_{\gam_1\cdots\gam_n}  = \frac{\delta^{(n)} \Phi[G]}{\delta
  G_{\gam_1} \cdots \delta G_{\gam_n}} .
\eeq 
To order $V^m$, $\Phi^{(n)}$ is represented by all 2PI diagrams with $n$
external (bosonic) legs $\gam_i$ and $2m-n$ internal lines. These diagrams
cannot be separated into two disconnected pieces by cutting two internal lines
(considering every external leg $\gam_i=\lbrace\alpha_i,\beta_i\rbrace$ as a
connected piece). The 2PI vertices satisfy the symmetry properties
\bleq
\Phi^{(n)}_{\gam_1\cdots\lbrace\alpha_i,\beta_i\rbrace\cdots\gam_n} &=& -
\Phi^{(n)}_{\gam_1\cdots\lbrace\beta_i,\alpha_i\rbrace\cdots\gam_n} ,
\nonumber \\ 
\Phi^{(n)}_{\gam_1\cdots\gamma_i\cdots\gam_j\cdots\gam_n} &=&
\Phi^{(n)}_{\gam_1\cdots\gamma_j\cdots\gam_i\cdots\gam_n} .
\label{phi_sym} 
\eleq
The equation of motion (\ref{state_eq}) can be rewritten as a Dyson equation 
\beq
G^{-1}_\gam = C^{-1}_\gam - J_\gam + \Sigma_\gam ,
\label{Sigma1}
\eeq
with $\Sigma_\gam = \Phi^{(1)}_\gam$ the self-energy. Note that the
diagrammatic interpretation of $\Phi[G]$ as the sum of 2PI vacuum fluctuation
diagrams follows from Eq.~(\ref{Sigma1}) (see, for instance,
Ref.~\onlinecite{Haussmann99}). Similarly, the
equation $\Gamma^{(2)} W^{(2)}=I$ can be recast in the form  
\bleq
W^{(2)}_{\gam_1\gam_2} &=& G_{\alpha_1\beta_2} G_{\beta_1\alpha_2} -
G_{\alpha_1\alpha_2} G_{\beta_1\beta_2} \nonumber \\ 
&& + \half \sum_{\gam_3,\gam_4} G_{\alpha_1\alpha_3} G_{\beta_1\beta_3}
\Phi^{(2)}_{\gam_3\gam_4} W^{(2)}_{\gam_4\gam_2} .
\label{bethe}
\eleq 
Eq.~(\ref{bethe}) is a Bethe-Salpeter equation relating the two-particle Green
function $W^{(2)}$ to the 2PI vertex $\Phi^{(2)}$.

Equilibrium quantities are obtained for vanishing external sources ($J=0$). 
The equilibrium Green function $\bar G$ is determined by the stationary
condition
\beq
\frac{\delta \Gamma[G]}{\delta G_\gam} \biggr|_{\bar G} = 0 .
\label{saddle}
\eeq
It can be expressed in terms of the self-energy $\bar\Sigma=\Sigma|_{\bar G} =
\bar\Phi^{(1)}$,  
\beq
\bar G^{-1}_\gam = C^{-1}_\gam + \bar\Sigma_\gam . 
\eeq
The thermodynamic potential is given by $\Omega=-\beta^{-1}\ln Z[J=0]
=\beta^{-1}\Gamma[\bar G]$.

\subsection{2PI RG equations}
\label{subsec:rgeq_gen}

We now take the free propagator $C$ to depend on an infrared cutoff $\Lambda$
which suppresses the low-energy degrees of freedom ($|\xi_\k|\lesssim
\Lambda$), 
\beq
C(\k,i\wn) = - \frac{\Theta_\k}{i\wn-\xi_\k} .
\label{C_def} 
\eeq
$C(\k,i\wn)$ is the Fourier transform of $C(\r,\tau,-;\r',\tau',+)$, which we
assume to be spin-rotation invariant. $\xi_{\k}=\eps_{\k}-\mu$ is the
dispersion of the free fermions (with $\mu$ the chemical potential), and $\wn$
a fermionic Matsubara frequency. $\Theta_\k$ is a cutoff function such that
$\Theta_\k|_{\Lambda=0}=1$ and $\Theta_\k|_{\Lambda=\Lambda_0}=0$, where
$\Lambda_0=\max_\k |\xi_\k|$. Physical quantities are obtained for
$\Lambda=0$, when all degrees of freedom are included in the partition
function. $W[J]$ and its Legendre transform $\Gamma[G]$ now depend on
$\Lambda$ and satisfy flow equations as the cutoff $\Lambda$ is varied between
$\Lambda_0$ and 0.  

From the definition of $\Gamma[G]$ [Eq.~(\ref{Gamma_def})], we deduce
\bleq
\frac{d}{dl} \Gamma[G] &=& - \frac{\partial}{\partial l} W[J] - \half
\sum_\gam \frac{dJ_\gam}{dl} \frac{\delta W[J]}{\delta J_\gam}
- \half \Tr\left( \frac{dJ}{dl} G \right)
\nonumber \\ 
&=& - \frac{\partial}{\partial l} W[J] ,
\label{rgeq1}
\eleq 
where $l$ can be either $\Lambda$ or a function of $\Lambda$,
e.g. $l=\ln(\Lambda_0/\Lambda)$. Recall that $J$ depends on
$G$ {\it via} Eq.~(\ref{G_def}) and is therefore a function of $l$. $\partial
W[J]/\partial l$ denotes the variation of $W[J]=\ln Z[J]$ which follows from
the explicit $\Lambda$ dependence of the propagator $C$, 
\bleq
\frac{\partial}{\partial l} W[J] &=& \frac{1}{Z[J]} \int \calD[\psi] \left(
-\half \psi^T \dot C^{-1} \psi \right) 
e^{-S[\psi]+\half\psi^T J \psi} \nonumber \\ 
&=& -\half \sum_{\alpha,\beta} \dot C^{-1}_{\alpha\beta} \mean{\psi_\alpha
  \psi_\beta} \nonumber \\ 
&=& \half \Tr (\dot C^{-1}G) ,
\label{rgeq2}
\eleq
where the dot denotes a derivation with respect to $l$. 
Eqs.~(\ref{rgeq1},\ref{rgeq2}) imply
\beq
\frac{d}{dl} \Gamma[G] = - \half \Tr(\dot C^{-1} G)  ,
\label{Gamma_rg}
\eeq 
and, using Eq.~(\ref{phi_def}), 
\beq
\frac{d}{dl} \Phi[G] = 0 .
\label{Phi_rg}
\eeq 
The LW functional is invariant under the RG transformation. It is therefore a
``universal'' functional independent of the free propagator $C$.\cite{note7} 
This property
has a simple diagrammatic interpretation. Being the sum of the 2PI graphs
(with the internal lines corresponding to the variable $G$), $\Phi[G]$ depends
on the interaction vertex $V$, but not on the non-interacting
($\Lambda$-dependent) propagator $C$. 

The thermodynamic potential satisfies the RG equation
\bleq
\dot\Omega &=& \frac{1}{\beta} \frac{d}{dl} \left( \Gamma[G]\Bigr|_{\bar G}
\right) \nonumber \\ 
&=& \frac{1}{\beta}  \left(\frac{d}{dl} \Gamma[G]\right)_{\bar G} +
\frac{1}{2\beta} \sum_\gam \frac{\delta \Gamma[G]}{\delta G_\gam}
\biggr|_{\bar G} \dot{\bar G}_\gam \nonumber \\ 
&=& - \frac{1}{2\beta} \Tr(\dot C^{-1} \bar G) ,
\label{Omega_rg}
\eleq 
where the last line is obtained using equations (\ref{Gamma_rg}) and
(\ref{saddle}).  

\begin{figure}
\includegraphics[width=3cm]{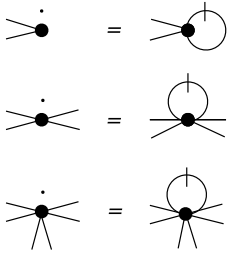}
\caption{Diagrammatic representation of the hierarchy (\ref{rg_eq1}) of flow
  equations satisfied by the 2PI vertices
  $\bar\Phi^{(n)}_{\gam_1\cdots\gam_n}$. Derivation with respect to $l$
  are indicated by dots (vertices) and slashed lines (propagators).
  A pair of neighboring external legs corresponds to
  $\gam_i=\lbrace\alpha_i,\beta_i\rbrace$. } 
\label{fig:rgeq_gen}
\end{figure}

Since $\Phi[G]$ is invariant under the RG transformation, the flow of the 2PI
vertices $\bar\Phi^{(n)}=\Phi^{(n)}|_{\bar G}$ is entirely due to the flow of
$\bar G$,
\bleq
\dot {\bar \Phi}^{(n)}_{\gam_1 \cdots \gam_n} &=& \frac{d}{dl} 
\left(\frac{\delta^{(n)} \Phi[G]}{\delta G_{\gam_1}\cdots \delta G_{\gam_n}} 
\biggr|_{\bar G} \right) \nonumber \\ 
&=& \half \sum_\gam \frac{\delta^{(n+1)} \Phi[G]}{\delta G_{\gam_1}\cdots
  \delta G_{\gam_n}\delta G_\gam} \biggr|_{\bar G} \frac{d}{dl} \bar G_\gam 
\nonumber \\ 
&=& \half \sum_\gam \bar\Phi^{(n+1)}_{\gam_1\cdots \gam_n\gam} 
\dot{\bar G}_\gam . 
\label{rg_eq1}
\eleq
We thus obtain an infinite hierarchy of flow equations for the 2PI vertices
(Fig.~\ref{fig:rgeq_gen}). In all Feynman diagrams shown in this paper, 
a pair of neighboring external legs
corresponds to a bosonic index $\gam_i=\lbrace\alpha_i,\beta_i\rbrace$. 
We shall always represent the 2PI two-particle vertex 
$\bar\Phi^{(2)}_{\gam_1\gam_2}$
with the two external legs $\gam_1=\lbrace\alpha_1,\beta_1\rbrace$
($\gam_2=\lbrace\alpha_2,\beta_2\rbrace$) on the left (right) hand side of the
vertex: $\bar\Phi^{(2)}_{\gam_1\gam_2}$ is 2PI as seen from left to right. 

Eqs.~(\ref{Omega_rg},\ref{rg_eq1}) should be supplemented with the
initial values of the thermodynamic potential and the 2PI vertices at
$\Lambda=\Lambda_0$. Since the cutoff function $\Theta_\k$ is chosen such that 
$C|_{\Lambda_0}=0$, we easily deduce from the diagrammatic expansion
of the LW functional that $\bar\Sigma|_{\Lambda_0}=0$ (i.e. $\bar
G|_{\Lambda_0}=0$), $\bar\Phi^{(2)}|_{\Lambda_0}=V$, and
$\bar\Phi^{(n)}|_{\Lambda_0}=0$ for $n\geq 3$.  
Using Eq.~(\ref{phi_def}), we also obtain $\Omega|_{\Lambda_0}
=\Omega_0|_{\Lambda_0}$ where $\Omega_0=(2\beta)^{-1}\Tr\ln C$ is the
non-interacting 
thermodynamic potential obtained from the action $S_0$. To avoid computing
$\Omega_0$, we shall always calculate $\Delta\Omega=\Omega-\Omega_0$. 

For practical calculations, one has to truncate the hierarchy of flow
equations (\ref{rg_eq1}) by retaining a
finite number of low-order vertices. The simplest non-trivial truncation is
discussed in the next section. Unless mentioned otherwise, we now drop the
``bar'' above Green functions and vertices since we shall only consider the
case $J=0$.

\subsection{One-loop RG equations}
\label{subsec:rgeq_1L}

\subsubsection{Flow equations for the 2PI vertices}

\begin{figure}
\includegraphics[width=7.5cm,bb=97 487 362 611]{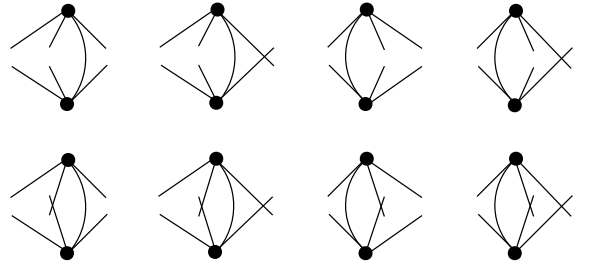}
\caption{Diagrammatic representation of $\Phi^{(3)}$ as a function of $\Phid$
  within the one-loop approximation. $\Phid$, shown as a black dot, is 2PI as
  seen from left to right. Signs and symmetry factors are not indicated. }
\label{fig:phi3}
\end{figure}

One-loop RG equations are obtained by neglecting $\Phi^{(n)}$ for $n\geq
4$. This reduces the hierarchy of equations (\ref{rg_eq1}) to
\bleq
\dot \Sigma_{\gam_1} &=& \half \sum_{\gam_2} \Phi^{(2)}_{\gam_1\gam_2}
\dot G_{\gam_2} , \nonumber \\
\dot \Phi^{(2)}_{\gam_1\gam_2} &=& \half \sum_{\gam_3}
\Phi^{(3)}_{\gam_1\gam_2\gam_3} \dot G_{\gam_3} .
\label{rg_eq2} 
\eleq
In order to close this system of equations, we need an approximate expression
of $\Phi^{(3)}$ in terms of $\Phid$ and $\Sigma$. Let us start with the
second-order contribution to the LW functional,
\beq
\Phi[G] = - \frac{1}{48} \sum_{\gam_1,\gam_2\gam_3,\gam_4}
V_{\alpha_1\beta_3\alpha_4\beta_2} V_{\alpha_3\beta_1\alpha_2\beta_4} 
G_{\gam_1} G_{\gam_2}G_{\gam_3} G_{\gam_4} .
\label{phi_1}
\eeq
By taking the third-order functional derivative with respect to $G$, we obtain
\bleq 
\Phi^{(3)}_{\gam_1\gam_2\gam_3} &=& - \half \sum_{\gam_4} G_{\gam_4} 
[ V_{\alpha_1\alpha_3\alpha_2\alpha_4} V_{\beta_1\beta_3\beta_2\beta_4} 
\nonumber \\ && -
V_{\alpha_1\alpha_3\beta_2\alpha_4} V_{\beta_1\beta_3\alpha_2\beta_4} +
V_{\alpha_1\beta_4\beta_2\alpha_3} V_{\beta_1\alpha_4\alpha_2\beta_3} 
\nonumber \\ && -
V_{\alpha_1\beta_4\alpha_2\alpha_3} V_{\beta_1\alpha_4\beta_2\beta_3} -
(\alpha_3 \leftrightarrow \beta_3) ] .
\label{phi3_1} 
\eleq 
Replacing $V$ by $\Phi^{(2)}$ in Eq.~(\ref{phi3_1}),\cite{Wetterich02} 
we obtain the one-loop approximation of $\Phi^{(3)}$, 
\begin{multline}
\Phi^{(3)}_{\gam_1\gam_2\gam_3}\Bigl|_{\rm 1\,loop} 
= - \half \sum_{\gam_4} G_{\gam_4} 
\bigl[ \Phi^{(2)}_{\alpha_1\alpha_3\alpha_2\alpha_4}
\Phi^{(2)}_{\beta_1\beta_3\beta_2\beta_4}  \\ -
\Phi^{(2)}_{\alpha_1\alpha_3\beta_2\alpha_4}
\Phi^{(2)}_{\beta_1\beta_3\alpha_2\beta_4} + 
\Phi^{(2)}_{\alpha_1\beta_4\beta_2\alpha_3}
\Phi^{(2)}_{\beta_1\alpha_4\alpha_2\beta_3}  \\ -
\Phi^{(2)}_{\alpha_1\beta_4\alpha_2\alpha_3}
\Phi^{(2)}_{\beta_1\alpha_4\beta_2\beta_3} - 
(\alpha_3 \leftrightarrow \beta_3) \bigr] .
\label{phi3_2}
\end{multline}
By working out the symmetry factors of various diagrammatic contributions to
$\Phid$ and $\Phi^{(3)}$, one can convince oneself that the overall factor
$-1/2$ in Eq.~(\ref{phi3_1}) remains unchanged when $V$ is replaced by
$\Phi^{(2)}$. $\Phi^{(3)}|_{\rm 1\,loop}$ is shown diagrammatically in
Fig.~\ref{fig:phi3}.  

\begin{figure}
\includegraphics[width=8cm,bb=93 445 385 732]{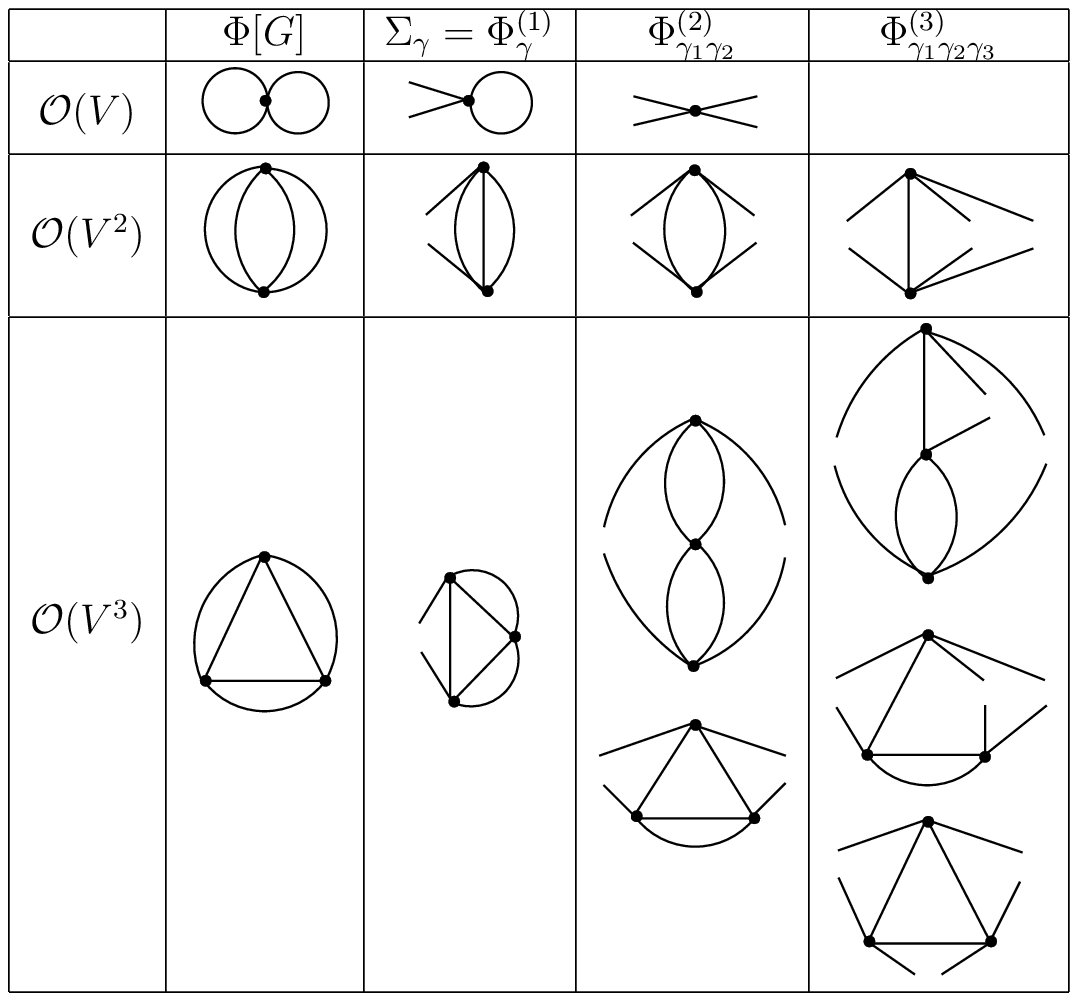}
\caption{Diagrams contributing to $\Phi[G]$, $\Sigma_{\gam_1}$,
  $\Phid_{\gam_1\gam_2}$ and $\Phi^{(3)}_{\gam_1\gam_2\gam_3}$ up to third
  order in the bare interaction vertex $V$
  (shown as a (small) black dot). Diagrams obtained by exchanging 
  external legs ($\alpha_i\leftrightarrow\beta_i$) are not shown. } 
\label{fig:V3}
\end{figure}

Diagrams contributing to $\Phi[G]$, $\Sigma$, $\Phid$ and $\Phi^{(3)}$ up to
third order in the bare interaction amplitude $V$ are shown in
Fig.~\ref{fig:V3}. The $\calO(V^2)$ contribution to $\Phi^{(3)}$ is included
in the one-loop approximation, but among the three diagrammatic contributions
to order $V^3$ only the first one is retained. The other two
are not of the form (\ref{phi3_2}); the second one involves a
two-particle-reducible two-particle vertex, while the third one 
is clearly not of the required type. Thus, a given diagram contributing to the
LW functional will generate diagrams for $\Phi^{(3)}$ which may or may not be
included in the one-loop approximation. The latter is therefore not a
$\Phi$-derivable approximation as it is not based on a truncation of the LW
functional. A detailed study of conservation laws and Ward identities in the
2PI RG scheme is beyond the scope of this work and remains to be done. 
   
It should also be noticed that the one-loop approximation, like any
approximation of the 2PI vertex $\Phid$, leads to a violation of the
crossing symmetries of the two-particle Green function $W^{(2)}$
(e.g. $W^{(2)}_{\alpha_1\beta_1\alpha_2\beta_2} \neq -
W^{(2)}_{\alpha_1\alpha_2\beta_1\beta_2}$) and the 1PI two-particle vertex.  

\begin{figure}
\includegraphics[width=6cm]{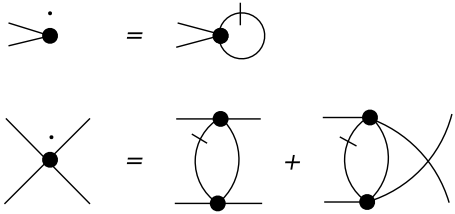}
\caption{One-loop flow equations for $\Sigma_\gam$ and $\Phid_{\gam_1\gam_2}$.
Here and in the following figures, diagrams obtained by exchanging the slashed
and non-slashed lines in the one-loop diagrams contributing to
${\dot\Phi}^{(2)}$ are not shown. } 
\label{fig:rgeq_1L}
\end{figure}

From Eqs.~(\ref{rg_eq2},\ref{phi3_2}), we finally obtain the one-loop RG
equations 
\bleq
\dot \Sigma_{\gam_1} &=& \half \sum_{\gam_2} \Phi^{(2)}_{\gam_1\gam_2} 
\dot G_{\gam_2} , \nonumber \\ 
\dot \Phi^{(2)}_{\gam_1\gam_2} &=& \half \sum_{\gam_3,\gam_4} (\dot G_{\gam_3}
G_{\gam_4} + G_{\gam_3} \dot G_{\gam_4} ) \nonumber \\ && \times
\bigl[ \Phi^{(2)}_{\alpha_1\alpha_3\alpha_4\beta_2} 
\Phi^{(2)}_{\beta_1\beta_3\beta_4\alpha_2}  -
\Phi^{(2)}_{\alpha_1\alpha_3\alpha_4\alpha_2} 
\Phi^{(2)}_{\beta_1\beta_3\beta_4\beta_2} \bigr] . 
\nonumber \\ && 
\label{rg_eq3}
\eleq  
Eqs.~(\ref{rg_eq3}) are shown diagrammatically in Fig.~\ref{fig:rgeq_1L}. 
There are two differences with respect to the one-loop RG equations obtained
within the 1PI RG scheme:\cite{Salmhofer01,Honerkamp01} (i) the flow equation
for the self-energy involves the 2PI two-particle vertex and the derivative
$\dot G$ of the Green function instead of the 1PI vertex and the
``single-scale propagator'' $S=-G\dot C^{-1}G$; (ii) the one-loop contribution
which would give a two-particle-reducible contribution to $\Phid$ is
absent.

The procedure we have followed to obtain the
one-loop RG equation is not unique. Owing to the antisymmetry of $V$, one
could write $-V_{\alpha_1\beta_3\alpha_4\beta_2}
V_{\alpha_3\beta_4\alpha_2\beta_1}$  or $V_{\alpha_4\beta_3\alpha_1\beta_2}
V_{\alpha_3\beta_4\alpha_2\beta_1}$ instead of
$V_{\alpha_1\beta_3\alpha_4\beta_2} V_{\alpha_3\beta_1\alpha_2\beta_4}$
in Eq.~(\ref{phi_1}). This explains while the one-loop RG equation obtained
in Ref.~\onlinecite{Wetterich02},
\begin{multline}
\dot \Phi^{(2)}_{\gam_1\gam_2} = \sum_{\gam_3,\gam_4} \dot G_{\gam_3}
G_{\gam_4} 
\bigl[ \Phi^{(2)}_{\alpha_4\alpha_3\alpha_1\beta_2} 
\Phi^{(2)}_{\beta_4\beta_3\beta_1\alpha_2}  \\  -
\Phi^{(2)}_{\alpha_4\alpha_3\alpha_1\alpha_2} 
\Phi^{(2)}_{\beta_4\beta_3\beta_1\beta_2} \bigr] ,
\label{rg_eq3bis}
\end{multline}
differs from ours. This equation can be represented diagrammatic as in 
Fig.~\ref{fig:rgeq_1L}, but with the vertices $\Phid$ in the one-loop
corrections being 2PI as seen from top to
bottom. Eqs.~(\ref{rg_eq3},\ref{rg_eq3bis}) lead to different diagram
resummations. While the two $\calO(V^3)$ contributions to $\Phid$ in
Fig.~\ref{fig:V3} are generated by the equations (\ref{rg_eq3}), the first one
is not if one uses Eq.~(\ref{rg_eq3bis}). More generally, similar diagrams with
an arbitrary number of loops are not included in Eq.~(\ref{rg_eq3bis}). These
diagrams play a crucial role in most applications of the RG approach to
interacting fermion systems. For instance, they describe the exchange of spin
fluctuations in a conductor with short-range antiferromagnetic order and may
lead to $d$-wave or other types of unconventional superconductivity. For
this reason, we do not expect the one-loop approximation based on
Eq.~(\ref{rg_eq3bis}) to give reliable results.

\subsubsection{Relation to the 1PI RG scheme}

In this section we show how, starting from Eqs.~(\ref{rg_eq3}), we can
reproduce the one-loop RG equations for the 1PI two-particle vertex
$\gamma^{(4)}$ obtained within the 1PI RG scheme. $\gamma^{(4)}$ is defined by
\beq
G^{(4)}_{c,\gam_1\gam_2} = -
\sum_{\gam_3,\gam_4} G_{\alpha_1\alpha_3} G_{\beta_1\beta_3}
G_{\alpha_2\alpha_4} G_{\beta_2\beta_4} 
\gamma^{(4)}_{\gam_3\gam_4} ,
\label{gamma4_def}
\eeq
where
\bleq
G^{(4)}_{c,\gam_1\gam_2} &=& \mean{\psi_{\alpha_1}
  \psi_{\beta_1} \psi_{\alpha_2} \psi_{\beta_2} }
- G_{\alpha_1\beta_1} G_{\alpha_2\beta_2} \nonumber \\ &&
+ G_{\alpha_1\alpha_2} G_{\beta_1\beta_2}  
- G_{\alpha_1\beta_2} G_{\beta_1\alpha_2} \nonumber \\ 
&=& W^{(2)}_{\gam_1\gam_2} -  G_{\alpha_1\beta_2} G_{\beta_1\alpha_2} 
+ G_{\alpha_1\alpha_2} G_{\beta_1\beta_2}
\eleq
is the (fully) connected two-particle Green function. The Bethe-Salpeter
equation (\ref{bethe}) implies that $G^{(4)}_c$ satisfies 
\begin{multline}
\half \sum_{\gam_4} \Bigl[ I_{\gam_1\gam_4} + \half \sum_{\gam_3}
  \Pi_{\gam_1\gam_3} \Phi^{(2)}_{\gam_3\gam_4} \Bigr] G^{(4)}_{c,\gam_4\gam_2}
= \\
- \quarter \sum_{\gam_3,\gam_4} \Pi_{\gam_1\gam_3} \Phi^{(2)}_{\gam_3\gam_4}
  \Pi_{\gam_4\gam_2} ,
\label{Gc4_1}
\end{multline}
where 
\beq
\Pi_{\gam_1\gam_2} = G_{\alpha_1\beta_2}G_{\beta_1\alpha_2} -
G_{\alpha_1\alpha_2} G_{\beta_1\beta_2} .
\eeq
In matrix form, Eqs.~(\ref{gamma4_def},\ref{Gc4_1})  read
\bleq
G^{(4)}_c &=& - \Pi \gamma^{(4)} \Pi , \nonumber \\ 
\bigl( I+\Pi\Phid \bigr )G^{(4)}_c &=& -\Pi\Phid\Pi .
\eleq
We therefore obtain
\bleq
\gamma^{(4)} &=& \Phid \bigl( I+\Pi \Phid \bigr)^{-1} , \nonumber \\
\Phid &=& \gamma^{(4)} \bigl( I-\Pi \gamma^{(4)} \bigr)^{-1} .
\label{gamma4_phi2} 
\eleq
From these two equations, we deduce the following flow equation for the 1PI
vertex $\gamma^{(4)}$,
\bleq
\dot \gamma^{(4)} &=& - \gamma^{(4)} \dot \Pi \gamma^{(4)} 
+ \gamma^{(4)} \bigl(\Phid\bigr)^{-1} {\dot\Phi}^{(2)} \bigl(\Phid\bigr)^{-1}
\gamma^{(4)} 
\nonumber \\
&=&  - \gamma^{(4)} \dot \Pi \gamma^{(4)}  
+ \bigl(I-\gamma^{(4)}\Pi\bigr) {\dot\Phi}^{(2)} \bigl(I-\Pi\gamma^{(4)}\bigr).
\label{gamma4_rg1}
\eleq 
We can now reproduce the one-loop RG equations derived
within the 1PI RG scheme by expanding the rhs of Eq.~(\ref{gamma4_rg1}) to
second order in $\gamma^{(4)}$. Since $\Phid=\gamma^{(4)} +
\calO[(\gamma^{(4)})^2]$ and ${\dot\Phi}^{(2)}=\calO[(\Phid)^2]$, 
Eq.~(\ref{gamma4_rg1}) gives $\dot\gamma^{(4)}=-\gamma^{(4)} \dot\Pi
\gamma^{(4)}+{\dot\Phi}^{(2)}$ where, to order $(\gamma^{(4)})^2$,
${\dot\Phi}^{(2)}$ is given by Eq.~(\ref{rg_eq3}) with $\Phid$ replaced by
$\gamma^{(4)}$ in the rhs. This eventually gives
\begin{multline}
\dot\gamma^{(4)}_{\gam_1\gam_2} = - \half \sum_{\gam_3,\gam_4} 
\bigl( \dot G_{\alpha_3\beta_4} G_{\beta_3\alpha_4} + 
       G_{\alpha_3\beta_4} \dot G_{\beta_3\alpha_4} \bigr) 
\\ \times \bigl[ 
\gamma^{(4)}_{\alpha_1\beta_1\alpha_3\beta_3}
\gamma^{(4)}_{\alpha_4\beta_4\alpha_2\beta_2} - 
\gamma^{(4)}_{\alpha_1\alpha_3\beta_3\beta_2}
\gamma^{(4)}_{\beta_1\beta_4\alpha_4\alpha_2} 
\\ +
\gamma^{(4)}_{\alpha_1\alpha_3\beta_3\alpha_2}
\gamma^{(4)}_{\beta_1\beta_4\alpha_4\beta_2}
\bigr] .
\label{rg1PI_1}
\end{multline} 
To lowest order in $\gamma^{(4)}$, the flow equation (\ref{rg_eq3}) for the
self-energy becomes
\beq
\dot\Sigma_{\gam_1} = \half \sum_{\gam_2} \gamma^{(4)}_{\gam_1\gam_2} \dot
G_{\gam_2} .
\label{rg1PI_2}
\eeq
Eqs.~(\ref{rg1PI_1},\ref{rg1PI_2}) agrees with the equations derived
within the 1PI RG scheme\cite{Honerkamp01,Salmhofer01} with the ``single-scale
propagator'' $S=-G\dot C^{-1} G$ replaced by $\dot G$.

\subsection{One-loop RG equations in the normal phase} 
\label{subsec:rgeq_N}

In this section, we consider the one-loop RG equations in the absence of broken
symmetry. We denote the position, time and spin indices by $X$ so that
$\alpha=(X,c)$, $\psi(X+)=\psi^*(X)$, $\psi(X-)=\psi(X)$, and $\int
dX=\int_0^\beta d\tau \int d^dr \sum_\sig$. The
single-particle Green function is then given by
\beq
G(X_1-,X_2+) = \mean{\psi(X_1)\psi^*(X_2)} \equiv G(X_1,X_2) ,
\eeq
with $G(X_1+,X_2-)=-G(X_2,X_1)$. $G(X_1c,X_2c)$ vanishes in the normal phase. 
$G$ satisfies the Dyson equation 
\begin{multline}
G(X_1,X_2) = C(X_1,X_2) \\ 
- \int dX_3dX_4 C(X_1,X_3)\Sigma(X_3,X_4)G(X_4,X_2)
\end{multline}
with
\beq
\Sigma(X_1,X_2) = \Sigma(X_1+,X_2-)
\eeq
the self-energy. In the normal phase, the 2PI two-particle vertex
$\Phid(X_1c_1,X_2c_2,X_3c_3,X_4c_4)$ vanishes if $\sum_i
c_i \neq 0$. Distinguishing between the particle-particle (pp) and
particle-hole (ph) channels, we introduce
\bleq
\Phipp(X_1,X_2,X_3,X_4) &=& \Phid(X_1+,X_2+,X_3-,X_4-) , \nonumber \\
\Phiph(X_1,X_2,X_3,X_4) &=& \Phid(X_1+,X_2-,X_3+,X_4-) . \nonumber \\ && 
\label{Phippph}
\eleq 
Note that in Eqs.~(\ref{Phippph}), we have singled out one of the two
ph channels. The 2PI vertex in the other ph channel
is related to $\Phiph$ by
\beq
\Phid(X_1+,X_2-,X_3-,X_4+) = - \Phiph(X_1,X_2,X_4,X_3) .
\eeq
Since the one-loop approximation conserves the crossing symmetry
$\Phid_{\gam_1\lbrace\alpha_2\beta_2\rbrace} = 
- \Phid_{\gam_1\lbrace\beta_2\alpha_2\rbrace}$ (as obvious from
Fig.~\ref{fig:rgeq_1L}), it is possible to consider a single ph
channel. The symmetry properties of $\Phid_{\gam_1\gam_2}$
[Eqs.~(\ref{phi_sym})] imply 
\bleq
\Phipp(X_1,X_2,X_3,X_4) &=& -\Phipp(X_2,X_1,X_3,X_4) \nonumber \\ 
                        &=& -\Phipp(X_1,X_2,X_4,X_3) , \nonumber \\
\Phiph(X_1,X_2,X_3,X_4) &=& \Phiph(X_3,X_4,X_1,X_2) . 
\label{sym2}
\eleq
The RG equation for the self-energy can be written as 
\bleq
\dot \Sigma(X_1,X_2) &=& \half \sum_{c=\pm} \int dX_3 dX_4 
\dot G(X_3c,X_4\bar c) 
\nonumber \\ && \times \Phid(X_1+,X_2-,X_3c,X_4\bar c) \nonumber \\ 
&=& - \int dX_3 dX_4 \dot G(X_4,X_3) 
\nonumber \\ && \times\Phiph(X_1,X_2,X_3,X_4) ,
\label{rg_eq_self}
\eleq 
where $\bar c=-c$. A similar calculation for the two-particle vertices
$\Phipp$ and $\Phiph$ yields
\begin{widetext}
\bleq
{\dot\Phi}^{(2)}_{\rm pp}(X_i) 
&=& - \half \int dX'_1 dX'_2 dX'_3 dX'_4
\bigl[ G(X'_2,X'_1) \dot G(X'_3,X'_4) + (G\leftrightarrow \dot G) \bigr] 
\bigl[ \Phipp(X_1,X'_1,X'_3,X_4) \Phiph(X_2,X'_2,X'_4,X_3) \nonumber \\ && 
+ \Phiph(X_1,X'_2,X'_4,X_4) \Phipp(X_2,X'_1,X'_3,X_3) 
- (X_3 \leftrightarrow X_4) \bigr] ,
\nonumber \\ 
{\dot\Phi}^{(2)}_{\rm ph}(X_i) 
&=& - \half \int dX'_1 dX'_2 dX'_3 dX'_4
\bigl[ G(X'_2,X'_1) \dot G(X'_3,X'_4) + (G\leftrightarrow \dot G) \bigr]
\bigl[
\Phipp(X_1,X'_1,X'_3,X_4) \Phipp(X'_4,X_3,X_2,X'_2) \nonumber \\ && + 
\Phiph(X_1,X'_2,X'_4,X_4) \Phiph(X'_1,X_2,X_3,X'_3) 
+  \Phiph(X_1,X'_2,X_3,X'_3) \Phiph(X'_1,X_2,X'_4,X_4) \bigr] ,
\label{rg_eq31} 
\eleq  
where we use the short-hand notation
$\Phid(X_i)=\Phid(X_1,X_2,X_3,X_4)$. Eqs.~(\ref{rg_eq31}) are shown
diagrammatically in Fig.~\ref{fig:rgeq_N}. 

\begin{figure}
\includegraphics[width=8cm]{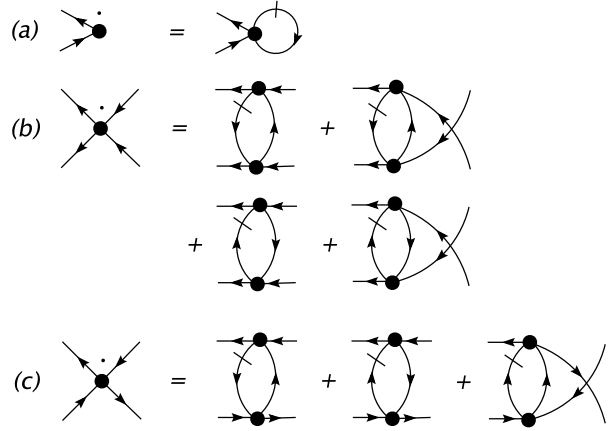}
\caption{One-loop RG equations for $\Sigma$ (a), $\Phipp$ (b) and $\Phiph$ (c)
  in the normal phase.} 
\label{fig:rgeq_N} 
\end{figure}

In spin-rotation invariant systems, it is convenient to write $\Phipp$ and
$\Phiph$ in the triplet/singlet and charge/spin basis, respectively,  
\bleq
\Phid_{{\rm pp},\sig_i}(x_i) &=& \Phit(x_i) I^{\sig_1\sig_2}_{\sig_3\sig_4} 
+ \Phis(x_i) T^{\sig_1\sig_2}_{\sig_3\sig_4} , \nonumber \\ 
\Phid_{{\rm ph},\sig_i}(x_i) &=& \Phich(x_i) \delta_{\sig_1,\sig_2}
\delta_{\sig_3,\sig_4} 
+  \Phisp(x_i) \bfsig_{\sig_1,\sig_2} \cdot \bfsig_{\sig_3,\sig_4} , 
\eleq
($X=(x,\sig)$, $x=(\r,\tau)$), where
\beq
I^{\sig_1\sig_2}_{\sig_3\sig_4} = \half \bigl(\delta_{\sig_1,\sig_4}
\delta_{\sig_2,\sig_3} + \delta_{\sig_1,\sig_3} \delta_{\sig_2,\sig_4} \bigr), 
\;\;\;\;\;
T^{\sig_1\sig_2}_{\sig_3\sig_4} = \half \bigl(\delta_{\sig_1,\sig_4}
\delta_{\sig_2,\sig_3} - \delta_{\sig_1,\sig_3} \delta_{\sig_2,\sig_4} \bigr),
\eeq
and $\bfsig=(\sig^x,\sig^y,\sig^z)$ stands for the Pauli matrices. From
Eqs.~(\ref{sym2}), we deduce the following symmetry properties
\bleq
\Phit(x_1,x_2,x_3,x_4) &=& - \Phit(x_2,x_1,x_3,x_4)
                       = - \Phit(x_1,x_2,x_4,x_3) , \nonumber \\
\Phis(x_1,x_2,x_3,x_4) &=& \Phis(x_2,x_1,x_3,x_4) 
                       = \Phis(x_1,x_2,x_4,x_3) , \nonumber \\
\Phid_{\rm ch,sp}(x_1,x_2,x_3,x_4) &=& \Phid_{\rm ch,sp}(x_3,x_4,x_1,x_2) .
\eleq
Performing the sum over spin indices in Eqs.~(\ref{rg_eq_self},\ref{rg_eq31}), 
we obtain the flow equations
satisfied by $\Sigma$, $\Phid_{\rm t,s}$ and $\Phid_{\rm ch,sp}$,
\beq
\dot\Sigma(x_1,x_2) = -2\int dx_3dx_4 \dot G(x_4,x_3) \Phich(x_1,x_2,x_3,x_4), 
\label{rg_eq4.0}
\eeq
\bleq
{\dot\Phi}^{(2)}_{\rm t}(x_i) 
&=& -\half \int dx'_1 dx'_2 dx'_3 dx'_4
B(x'_2,x'_1,x'_3,x'_4) \nonumber \\ && \times
\Bigl[ \Bigl( \Phit(x_1,x'_1,x'_3,x_4) \Phich(x_2,x'_2,x'_4,x_3) 
+ 2 \Phit \Phisp + \Phis \Phisp \Bigl) \nonumber \\ && 
+  \Bigl( \Phich(x_1,x'_2,x'_4,x_4) \Phit(x_2,x'_1,x'_3,x_3) +
2 \Phisp \Phit + \Phisp \Phis \Bigr)  
- (x_3 \leftrightarrow x_4) \Bigr] , \nonumber \\ 
{\dot\Phi}^{(2)}_{\rm s}(x_i) 
&=& -\half \int dx'_1 dx'_2 dx'_3 dx'_4
B(x'_2,x'_1,x'_3,x'_4) \nonumber \\ && \times
\Bigl[ \Bigl( \Phis(x_1,x'_1,x'_3,x_4) \Phich(x_2,x'_2,x'_4,x_3) 
+ 3 \Phit \Phisp \Bigr) \nonumber \\ && +
\Phich(x_1,x'_2,x'_4,x_4) \Phis(x_2,x'_1,x'_3,x_3) +
3 \Phisp \Phit \Bigr)  
+ (x_3 \leftrightarrow x_4) \Bigr] , \nonumber \\ 
{\dot\Phi}^{(2)}_{\rm ch}(x_i) 
&=& -\half \int dx'_1 dx'_2 dx'_3 dx'_4
B(x'_2,x'_1,x'_3,x'_4) \nonumber \\ && \times \biggl[ \biggl(
\frac{3}{4} \Phit(x_1,x'_1,x'_3,x_4) \Phit(x'_4,x_3,x_2,x'_2) +
\quarter \Phis \Phis \biggr) \nonumber \\ && + \Bigl(
\Phich(x_1,x'_2,x'_4,x_4) \Phich(x'_1,x_2,x_3,x'_3) +
3 \Phisp \Phisp \Bigr) \nonumber \\ && + \Bigl(
\Phich(x_1,x'_2,x_3,x'_3) \Phich(x'_1,x_2,x'_4,x_4) +
3 \Phisp \Phisp \Bigr) \biggr] ,  \nonumber \\ 
{\dot\Phi}^{(2)}_{\rm sp}(x_i) 
&=& -\half \int dx'_1 dx'_2 dx'_3 dx'_4
B(x'_2,x'_1,x'_3,x'_4) \nonumber \\ && \times \biggl[ \biggl(
\half \Phit(x_1,x'_1,x'_3,x_4) \Phit(x'_4,x_3,x_2,x'_2) +
\quarter \Phit \Phis +
\quarter \Phis \Phit \biggr) \nonumber \\ && + \Bigl(
\Phich(x_1,x'_2,x'_4,x_4) \Phisp(x'_1,x_2,x_3,x'_3) +
\Phisp \Phich + 2 \Phisp \Phisp \Bigr) \nonumber \\ && + \Bigl(
\Phich(x_1,x'_2,x_3,x'_3) \Phisp(x'_1,x_2,x'_4,x_4) +
\Phisp \Phich  - 2 \Phisp \Phisp \Bigr) \biggr] ,
\label{rg_eq4} 
\eleq
where 
\beq 
B(x_1,x_2,x_3,x_4) = G(x_1,x_2) \dot G(x_3,x_4) + (G\leftrightarrow \dot G) .
\eeq 
Due to spin-rotation invariance, the single-particle Green function
$G_{\sig_1\sig_2}(x_1,x_2)= \delta_{\sig_1,\sig_2}G(x_1,x_2)$ and the
self-energy $\Sigma_{\sig_1\sig_2}(x_1,x_2)= \delta_{\sig_1,\sig_2}
\Sigma(x_1,x_2)$. 
In Eqs.~(\ref{rg_eq4}), we have grouped inside parenthesis terms
with identical dependences on the variables $x_i,x'_i$ and thus avoided
unnecessary repetitions of the latter.

\subsection{Response functions in the normal phase}
\label{subsec:RFN}

Once we have solved the RG equations (\ref{rg_eq4.0},\ref{rg_eq4}) and
determined the 
self-energy $\Sigma$ and the 2PI vertex $\Phid$, we can obtain the
two-particle Green functions $W^{(2)}$ and the response functions from the
Bethe-Salpeter equation (\ref{bethe}). 

We define triplet- and singlet-pairing fields, and charge- and
spin-density fields by
\bleq
O^\nu_{\rm t}(x_1,x_2) &=& \left\lbrace
\begin{array}{lcl} 
\psi_\up(x_1) \psi_\up(x_2) & {\rm if} & \nu=1 , \\
\frac{1}{\sqrt{2}} \sum_\sig \psi_\sig(x_1) \psi_{\bar\sig}(x_2) & {\rm if} &
\nu=0 , \\ 
\psi_\down(x_1) \psi_\down(x_2) & {\rm if} & \nu=-1 ,
\end{array}
\right.  \nonumber \\ 
O_{\rm s}(x_1,x_2) &=& \frac{1}{\sqrt{2}} \sum_\sig \sig \psi_\sig(x_1)
\psi_{\bar\sig}(x_2) , \nonumber \\ 
\rho(x_1,x_2) &=& \sum_\sig \psi^*_\sig(x_1) \psi_\sig(x_2) , \nonumber \\ 
\S(x_1,x_2) &=& \sum_{\sig_1,\sig_2} \psi^*_{\sig_1}(x_1)
\bfsig_{\sig_1,\sig_2} \psi_{\sig_2}(x_2) , 
\eleq 
($\bar\sig=-\sig$), and the corresponding response functions
\bleq
\chi_{\rm t}(x_i) &=& \mean{ O^\nu_{\rm t}(x_1,x_2) O^{\nu*}_{\rm t}(x_4,x_3)}
= \left\lbrace 
\begin{array}{lcl}
W^{(2)}_{{\rm pp},\up\up\up\up}(x_i) & {\rm if} & \nu=1, \\
\half \sum_{\sig_1,\sig_3} 
W^{(2)}_{{\rm pp},\sig_1\bar\sig_1\sig_3\bar\sig_3}(x_i) &
      {\rm if} & \nu=0, \\
W^{(2)}_{{\rm pp},\down\down\down\down}(x_i) & {\rm if} & \nu=-1, 
\end{array}
\right. 
\nonumber \\ 
\chi_{\rm s}(x_i) &=& \mean{ O_{\rm s}(x_1,x_2) O^*_{\rm s}(x_4,x_3)} 
= \half \sum_{\sig_1,\sig_3} \sig_1 \bar\sig_3 
W^{(2)}_{{\rm pp},\sig_1\bar\sig_1\sig_3\bar\sig_3}(x_i) , \nonumber \\ 
\chi_{\rm ch}(x_i) &=& \mean{\rho(x_1,x_2) \rho(x_3,x_4)} 
= \sum_{\sig_1,\sig_3} W^{(2)}_{{\rm
    ph},\sig_1\sig_1\sig_3\sig_3}(x_2,x_1,x_4,x_3) , \nonumber \\ 
\chi_{\rm sp}(x_i) &=& \mean{S^\nu(x_1,x_2) S^\nu(x_3,x_4)} 
= \sum_{\sig_1,\sig_2,\sig_3,\sig_4} \sig^\nu_{\sig_1,\sig_2}
\sig^\nu_{\sig_3,\sig_4}  
W^{(2)}_{{\rm ph},\sig_2\sig_1\sig_4\sig_3}(x_2,x_1,x_4,x_3),   
\eleq 
where $W^{(2)}_{{\rm pp},\sig_i}(x_i)=W^{(2)}_{\sig_i}(x_1-,x_2-,x_3+,x_4+)$
and $W^{(2)}_{{\rm ph},\sig_i}(x_i)=W^{(2)}_{\sig_i}(x_1-,x_2+,x_3-,x_4+)$
are the two-particle Green functions in the pp and
ph channels, respectively. They satisfy the Bethe-Salpeter
equations [see Eq.~(\ref{bethe})] 
\bleq
W^{(2)}_{{\rm pp},\sig_i}(x_i) &=& \delta_{\sig_1,\sig_4}
\delta_{\sig_2,\sig_3} G(x_1,x_4) G(x_2,x_3) - \delta_{\sig_1,\sig_3}
\delta_{\sig_2,\sig_4} G(x_1,x_3) G(x_2,x_4) \nonumber \\ && 
+ \half \sum_{\sig,\sig'} \int dx'_1 dx'_2 dx'_3 dx'_4 G(x_1,x'_1)
G(x_2,x'_2) \Phid_{{\rm pp},\sig_1\sig_2\sig\sig'}(x'_1,x'_2,x'_3,x'_4) 
W^{(2)}_{{\rm pp},\sig\sig'\sig_3\sig_4}(x'_3,x'_4,x_3,x_4) ,
\nonumber \\ 
W^{(2)}_{{\rm ph},\sig_i}(x_i) &=& - \delta_{\sig_1,\sig_4}
\delta_{\sig_2,\sig_3} G(x_1,x_4) G(x_3,x_2) \nonumber \\ && 
+ \sum_{\sig,\sig'} \int dx'_1 dx'_2 dx'_3 dx'_4 G(x_1,x'_1)
G(x'_2,x_2) \Phid_{{\rm ph},\sig_1\sig_2\sig\sig'}(x'_1,x'_2,x'_3,x'_4) 
W^{(2)}_{{\rm ph},\sig'\sig\sig_3\sig_4}(x'_4,x'_3,x_3,x_4) .
\label{bethe_W2}
\eleq 
From Eqs.~(\ref{bethe_W2}) we deduce
\bleq
\chi_{\mu={\rm t,s}}(x_i) &=& G(x_1,x_4) G(x_2,x_3) \mp G(x_1,x_3) G(x_2,x_4) 
\nonumber \\ && 
- \half \int dx'_1 dx'_2 dx'_3 dx'_4 G(x_1,x'_1) G(x_2,x'_2)
\Phid_{\mu}(x'_1,x'_2,x'_3,x'_4) \chi_\mu(x'_4,x'_3,x_3,x_4) , 
\nonumber \\ 
\chi_{\mu={\rm ch,sp}}(x_i) &=& -2 G(x_4,x_1) G(x_2,x_3)
\nonumber \\ && 
+ 2 \int dx'_1 dx'_2 dx'_3 dx'_4 G(x_2,x'_1) G(x'_2,x_1)
\Phid_\mu(x'_1,x'_2,x'_3,x'_4) \chi_\mu(x'_3,x'_4,x_3,x_4) .
\label{chi}
\eleq
\end{widetext}
Eqs.~(\ref{chi}) enable ones to determine the response functions from the
knowledge of the single-particle Green function $G$ and the 2PI vertex
$\Phid$. 

In many cases, useful information can also be drawn from the 1PI two-particle
vertex 
$\gamma^{(4)}$. Rewriting the relation (\ref{gamma4_phi2}) between
$\gamma^{(4)}$ and $\Phid$ as
\beq
\gamma^{(4)}_{\gam_1\gam_2} = \Phid_{\gam_1\gam_2} - \half \sum_{\gam_3\gam_4}
\Phid_{\gam_1\gam_3} G_{\alpha_3\beta_4} G_{\beta_3\alpha_4}
\gamma^{(4)}_{\gam_4\gam_2} ,
\eeq
and considering this equation in the pp and ph
channels, we obtain
\bleq
\gamma^{(4)}_{\mu={\rm t,s}}(x_i) &=& \Phid_\mu(x_i) - \half \int dx'_1 dx'_2
dx'_3 dx'_4 \nonumber \\ && \times 
\Phid_\mu(x_1,x_2,x'_1,x'_2) G(x'_1,x'_3) G(x'_2,x'_4) \nonumber \\ && \times 
\gamma^{(4)}_\mu(x'_4,x'_3,x_3,x_4) , \nonumber \\ 
\gamma^{(4)}_{\mu={\rm ch,sp}}(x_i) &=& \Phid_\mu(x_i) + 2 \int dx'_1 dx'_2
dx'_3 dx'_4 \nonumber \\ && \times 
\Phid_\mu(x_1,x_2,x'_1,x'_2) G(x'_3,x'_1) G(x'_2,x'_4) \nonumber \\ && \times 
\gamma^{(4)}_\mu(x'_4,x'_3,x_3,x_4) .
\label{bethe_gamma4} 
\eleq

\subsection{Ginzburg-Landau expansion}
\label{subsec:GL} 

An essential feature of the 2PI scheme is the possibility
to continue the RG flow in a broken-symmetry phase. This will be illustrated in
the next section in the framework of the BCS theory, and further discussed in
the concluding section. As a byproduct, one can also derive
the Ginzburg-Landau expansion of the thermodynamical potential in the vicinity
of a phase transition. The interest of such an approach is that only the
solution of the RG equations with no symmetry breaking is necessary. When
several instabilities compete at low temperature, the derivation of the
Ginzburg-Landau expansion is expected to be much simpler than the full
solution of the RG equations in the broken-symmetry phase. 

We introduce the Legendre transform $F[\Sigma]$ of the LW functional
$\Phi[G]$,
\beq
F[\Sigma] = \Phi[G] + \half \Tr(G\Sigma) , 
\eeq
where $G\equiv G[\Sigma]$ is a functional of $\Sigma$ obtained by inverting
$\Sigma_\gam = \delta \Phi[G]/\delta G_\gam$. 
Here and in the following, we assume vanishing external
sources ($J=0$), and denote $\bar G$, $\bar\Sigma$ by $G$, $\Sigma$,
etc. $F[\Sigma]$ satisfies 
\beq
\frac{\delta F[\Sigma]}{\delta \Sigma_\gam} = - G_\gam .
\label{Fstate}
\eeq
The functional $F[\Sigma]$ allows us to rewrite $\Gamma[G]$ as a functional of
the self-energy,\cite{note2} 
\bleq
\Gamma[\Sigma] &=& - \half \Tr\ln(C^{-1}+\Sigma) + \half \Tr(G\Sigma) +
\Phi[G] \nonumber \\ 
&=& - \half \Tr\ln(C^{-1}+\Sigma) + F[\Sigma] ,
\label{gamma_sig}
\eleq
which is stationary at the equilibrium self-energy
$\Sigma\equiv\bar\Sigma$,
\beq
\frac{\delta \Gamma[\Sigma]}{\delta \Sigma_\gam} = 0 .
\label{eq_state2}
\eeq 

We now write the self-energy $\Sigma=\Sigma_N+\Delta$ as the sum of a normal
part $\Sigma_N$ and an anomalous part $\Delta$ which violates some
symmetries of the normal phase. $\Delta$ is an order parameter for the phase
transition. It can include different kinds of symmetry breaking,
such as antiferromagnetism and superconductivity. If we are able to solve the
flow equations in the normal phase, i.e. fixing $\Delta=0$, we can consider 
$\Gamma[\Delta]\equiv\Gamma[\Sigma_N+\Delta]$ as a functional of the (unknown)
anomalous self-energy $\Delta$. The latter is then determined from the
stationary condition (\ref{eq_state2}), 
\beq
\frac{\delta \Gamma[\Delta]}{\delta \Delta_\gam} = 0 .
\label{delta_stat} 
\eeq 
A crucial point here is that the 2PI flow equations with $\Delta=0$ can be
continued below the actual phase transition temperature $T_c$, since the 2PI
vertices do not become singular at the transition. The functional
$\Gamma[\Delta]$ can therefore be determined below $T_c$, where the
stationary value of $\Delta$ [Eq.~(\ref{delta_stat})] is finite. 

In the vicinity of the phase transition, where $\Delta$ is small, we expand
$\Gamma[\Delta]$ to fourth order,
\bleq 
\Gamma[\Delta] &=& \Gamma_N + \quarter \Tr(G_N\Delta)^2 
+ \frac{1}{2!}\frac{1}{2^2} \sum_{\gam_1,\gam_2} F^{(2)}_{\gam_1\gam_2}
  \Delta_{\gam_1} \Delta_{\gam_2} \nonumber \\ &&
- \frac{1}{6}  \Tr(G_N\Delta)^3 + \frac{1}{3!}\frac{1}{2^3}
\sum_{\gam_1,\gam_2,\gam_3} F^{(3)}_{\gam_1\gam_2\gam_3}
\Delta_{\gam_1} \Delta_{\gam_2} \Delta_{\gam_3} \nonumber \\  &&
+ \frac{1}{8}  \Tr(G_N\Delta)^4  + \frac{1}{4!}\frac{1}{2^4}
\sum_{\gam_1,\gam_2,\gam_3,\gam_4} F^{(4)}_{\gam_1\gam_2\gam_3\gam_4} 
\nonumber \\  && \times
\Delta_{\gam_1} \Delta_{\gam_2} \Delta_{\gam_3} \Delta_{\gam_4} ,
\label{GL} 
\eleq
where $\beta^{-1}\Gamma_N=\beta^{-1}\Gamma[\Sigma_N]=\Omega_N$ is the
thermodynamic potential and $G_N$ the Green function in the normal phase, and
\beq
F^{(n)}_{\gam_1\cdots\gam_n} = \frac{\delta^{(n)} F[\Sigma]}{\delta
  \Sigma_{\gam_1} \cdots \delta \Sigma_{\gam_n}} \biggr|_{\Sigma_N} .
\eeq
If, as in most one-loop approximations, one ignores the normal phase
self-energy ($G_N=C$), the evaluation of the terms $\Tr(G_N\Delta)^n$ does not
raise any difficulty.  
The determination of $\Gamma[\Delta]$ then requires the calculation of
the coefficients $F^{(n)}$. These can be related to the 2PI vertices
$\Phi^{(n)}$ by taking functional derivatives of Eq.~(\ref{Fstate}) with
respect to $G$. The first-order derivative gives
\beq
- \half \sum_{\gam_3} \frac{\delta^{(2)}\Phi[G]}{\delta G_{\gam_1} \delta
  G_{\gam_3}} \biggr|_{G_N} \frac{\delta^{(2)} F[\Sigma]}{\delta
  \Sigma_{\gam_3} \delta\Sigma_{\gam_2}} \biggl|_{\Sigma_N} 
  = I_{\gam_1\gam_2} ,
\eeq
i.e.
\beq
\bigl(F^{(2)}\bigr)^{-1}_{\gam_1\gam_2} = - \Phid_{\gam_1\gam_2} .
\eeq
Higher-order derivatives yield
\bleq
F^{(3)}_{\gam_1\gam_2\gam_3} &=& - \frac{1}{8} \sum_{\gam'_1,\gam'_2,\gam'_3}
\Phi^{(3)}_{\gam'_1\gam'_2\gam'_3} F^{(2)}_{\gam'_1\gam_1}
  F^{(2)}_{\gam'_2\gam_2} F^{(2)}_{\gam'_3\gam_3} , \nonumber \\  
F^{(4)}_{\gam_1\gam_2\gam_3\gam_4} &=&  \frac{1}{16}
\sum_{\gam'_1,\gam'_2,\gam'_3,\gam'_4} 
\Phi^{(4)}_{\gam'_1\gam'_2\gam'_3\gam'_4} F^{(2)}_{\gam'_1\gam_1}
  F^{(2)}_{\gam'_2\gam_2} F^{(2)}_{\gam'_3\gam_3} F^{(2)}_{\gam'_4\gam_4}
\nonumber \\ &&
- \frac{1}{8}  \sum_{\gam'_1,\gam'_2,\gam'_3}
\Phi^{(3)}_{\gam'_1\gam'_2\gam'_3} \bigl[ F^{(3)}_{\gam_4\gam'_1\gam_1}
  F^{(2)}_{\gam'_2\gam_2}  F^{(2)}_{\gam'_3\gam_3} \nonumber \\ &&
+ F^{(2)}_{\gam'_1\gam_1}  F^{(3)}_{\gam_4\gam'_2\gam_2}
F^{(2)}_{\gam'_3\gam_3} +  F^{(2)}_{\gam'_1\gam_1} F^{(2)}_{\gam'_2\gam_2}
F^{(3)}_{\gam_4\gam'_3\gam_3} \bigr] . \nonumber \\ && 
\eleq 
These equations simplify within the one-loop approximation, since $\Phi^{(n)}$
vanishes for $n\geq 4$ and $\Phi^{(3)}$ can be expressed in terms of
$\Phi^{(2)}$. $F^{(3)}$ and $F^{(4)}$ are then essentially determined by
$\Phid$ and $F^{(2)}$. Writing the equation $\Phid F^{(2)}=-I$ in the
pp and ph channels, we obtain
\begin{multline}
\half \int dXdX' \Phipp(X_1,X_2,X,X') F^{(2)}_{\rm pp} (X',X,X_3,X_4) =
\\
\shoveright{-\delta(X_1-X_4) \delta(X_2-X_3) 
            + \delta(X_1-X_3) \delta(X_2-X_4),}
\\ 
\shoveleft{\int dXdX' \Phiph(X_1,X_2,X,X') F^{(2)}_{\rm ph} (X',X,X_3,X_4) =}
\\
- \delta(X_1-X_4) \delta(X_2-X_3),
\end{multline} 
where 
\bleq F^{(2)}_{\rm pp}(X_1,X_2,X_3,X_4) &=& F^{(2)}(X_1-,X_2-,X_3+,X_4+) ,
\nonumber \\  
F^{(2)}_{\rm ph}(X_1,X_2,X_3,X_4) &=& F^{(2)}(X_1-,X_2+,X_3-,X_4+) .
\nonumber \\ && 
\eleq 
For a spin-rotation invariant system, we finally deduce
\begin{multline}
{\int dxdx' \Phit(x_1,x_2,x,x') F^{(2)}_{\rm t}
  (x',x,x_3,x_4)=}  \\
\shoveright{-2[\delta(x_1-x_4) \delta(x_2-x_3) 
- \delta(x_1-x_3) \delta(x_2-x_4)],} 
\\ 
\shoveleft{\int dxdx' \Phis(x_1,x_2,x,x') F^{(2)}_{\rm s} (x',x,x_3,x_4)
  =} \\
\shoveright{-2[\delta(x_1-x_4) \delta(x_2-x_3) 
+ \delta(x_1-x_3) \delta(x_2-x_4)],} 
\\
\shoveleft{\int dxdx' \Phich(x_1,x_2,x,x') F^{(2)}_{\rm ch} (x',x,x_3,x_4) =} 
\\
\shoveright{- \quarter \delta(x_1-x_4) \delta(x_2-x_3),}
\\
\shoveleft{\int dxdx' \Phisp(x_1,x_2,x,x') F^{(2)}_{\rm sp} (x',x,x_3,x_4) =} 
\\
- \quarter \delta(x_1-x_4) \delta(x_2-x_3) .
\label{Phi_F}
\end{multline}
where $F^{(2)}_{\rm t,s}$ are the triplet and singlet parts of $F^{(2)}_{\rm
pp}$, and $F^{(2)}_{\rm ch,sp}$ the charge and spin parts of $F^{(2)}_{\rm
ph}$. In the next section, we shall use Eq.~(\ref{GL}) to reproduce the
Ginzburg-Landau expansion of the thermodynamic potential of a BCS
superconductor. 

\section{BCS theory}
\label{sec:BCS}

The aim of this section is to reproduce the main results of the BCS theory
using the 2PI RG equations. We consider a 3D system described by the action
\bleq
S &=& \int dx \, \sum_\sig \psi^*_\sig(x) \left(\dtau -\mu-
\frac{\bfnabla^2_\r}{2m} \right) \psi_\sig(x) \nonumber \\ &&
+ \lambda \int dx \, \psi^*_\up(x)\psi^*_\down(x)\psi_\down(x)
\psi_\up(x), 
\eleq
where $m$ is the fermion mass, $\mu$ the chemical potential, and $\lambda<0$
the amplitude of the local attractive interaction. This singular interaction is
regularized by means of an ultraviolet cutoff acting on the fermion
dispersion: $|\xi_\k|<\Lambda_0$, $\xi_\k=\k^2/2m-\mu$. 

\subsection{BCS gap equation} 

\begin{figure}
\includegraphics[width=2cm]{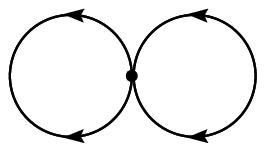}
\caption{Luttinger-Ward functional $\Phi_{\rm BCS}[G]$ in the BCS theory.}
\label{fig:Phi_BCS}
\end{figure}

There are two equivalent ways to derive the RG equations for the BCS
theory. One can start from the one-loop equations and neglect the ph
channel. Since the one-loop contribution to $\dot\Phi^{(2)}_{\rm pp}$ involves
only the ph channel, it vanishes in the BCS approximation, i.e. 
\beq
\dot\Phi^{(2)}_{\rm pp}=0 .
\label{phid_bcs}
\eeq 
Alternatively, one can start directly from
the LW functional (Fig.~\ref{fig:Phi_BCS})
\beq
\Phi_{\rm BCS}[G] = \lambda \int dx\, G_{\up\down}(x+,x+) G_{\down\up}(x-,x-) ,
\label{phi_bcs}
\eeq
where
\bleq
G_{\up\down}(x+,x'+) &=& \mean{\psi^*_\up(x)
  \psi^*_\down(x')} , \nonumber \\
G_{\down\up}(x-,x'-) &=& \mean{\psi_\down(x)
  \psi_\up(x')} .
\eleq
One then obtain $\Phi^{(3)}_{\gam_1\gam_2\gam_3}= 0$,
which leads to Eq.~(\ref{phid_bcs}). 

In the case of a local interaction, the only non-vanishing part of $\Phipp$
reads   
\bleq
\Phi^{(2)}_{{\rm pp},\sig_i}(x_i) &=& \tilde \Phi^{(2)}_{{\rm pp},\sig_i} 
\delta(x_1-x_2) \delta(x_2-x_3) \delta(x_3-x_4) \nonumber \\ 
\tilde \Phi^{(2)}_{{\rm pp},\up\down\down\up} &=& - \tilde \Phi^{(2)}_{{\rm
pp},\up\down\up\down} = - \tilde\Phi^{(2)}_{{\rm pp},\down\up\down\up} 
= \tilde\Phi^{(2)}_{{\rm pp},\down\up\up\down} = \lambda , \nonumber \\ &&
\label{Phi2_bcs}
\eleq
i.e. $\tilde\Phi^{(2)}_{\rm s} = 2\lambda$ and $\tilde\Phi^{(2)}_{\rm t}=0$. 
The self-energy has two non-vanishing elements, $\Sigma_{\up\down}(x+,y+)=
-\Sigma_{\down\up}(y+,x+)$ and $\Sigma_{\down\up}(x-,y-)= 
-\Sigma_{\up\down}(y-,x-)$, determined by the RG equations 
\bleq
{\dot\Sigma}_{\up\down}(x+,y+) &=& \lambda \delta(x-y) \dot
G_{\down\up}(x-,y-) , \nonumber \\ 
{\dot\Sigma}_{\down\up}(x-,y-) &=& \lambda \delta(x-y) \dot
G_{\up\down}(x+,y+) .
\eleq
Defining the superconducting order parameter $\Delta$ by 
\bleq
\Sigma_{\up\down}(x+,y+) &=& \delta(x-y) \Delta , \nonumber \\ 
\Sigma_{\down\up}(x-,y-) &=& \delta(x-y) \Delta^* , 
\eleq
the flow equation becomes
\beq
\dot\Delta = \lambda \dot G_{\down\up}(x-,x-) .
\label{dotdelta}
\eeq
Integrating Eq.~(\ref{dotdelta}) between $\Lambda_0$ and $\Lambda$, we obtain
the gap equation 
\beq
\Delta = \frac{\lambda}{\beta V} \sum_k G_{\down\up}(k-,k-) ,
\label{gapeq1}
\eeq
where $V$ is the volume of the system, and $k=(\k,i\wn)$ with $\wn$ a
fermionic Matsubara frequency. 

We use the standard Nambu notation, $\Psi(x)=(\psi_\up(x),\psi^*_\down(x))^T$,
to write the inverse Green function as a $2\times 2$ matrix in reciprocal
space, 
\beq
\calG^{-1}(k) = \left(
\begin{array}{cc} 
C^{-1}(k) & \Delta \\ 
\Delta^* & - C^{-1}(-k) 
\end{array}
\right) ,
\label{Ginverse}
\eeq
where $C(k)$ is the free propagator defined by Eq.~(\ref{C_def}). For 
intermediate calculations, it is convenient to assume that $\Theta_\k$ is a
smooth cutoff function which does not vanish ($\Theta_\k\neq 0$). Final
results depend only on $C(k)$ (and not $C^{-1}(k)$) and are well defined even
for a hard cutoff $\Theta_\k=\Theta(|\xi_\k|-\Lambda)$ ($\Theta$ is the step
function). Inverting $\calG^{-1}$, we obtain 
\bleq
G_{\sig\sig}(k-,k+) = \Theta_\k \frac{i\wn+\xik}{\wn^2+\Ek^2} ,
\nonumber \\ 
G_{\up\down}(k-,k-) = \Theta_\k \frac{\Delta_\k}{\wn^2+\Ek^2} , 
\nonumber \\ 
G_{\down\up}(k+,k+) = \Theta_\k
\frac{\Delta^*_\k}{\wn^2+\Ek^2} ,
\label{GF_def}
\eleq 
where 
\beq
\Ek = \sqrt{\xik^2 + |\Delta_\k^2|}, \;\;\;\; \Delta_\k = \Delta \Theta_\k .
\eeq 
Using Eqs.~(\ref{GF_def}), we can rewrite the gap equation (\ref{gapeq1}) as 
\beq
\Delta = - \frac{\lambda}{V} \sum_\k \frac{\Theta^2_\k\Delta}{2\Ek}
\tanh\left(\beta \frac{\Ek}{2}\right) .
\label{gapeq11}
\eeq
When $\Delta\neq 0$, this equation becomes
\beq
\frac{1}{|\tilde\lambda|} = \int_\Lambda^{\Lambda_0}
\frac{d\xi}{\sqrt{\xi^2+|\Delta|^2}} \tanh \left( \frac{\beta}{2}
\sqrt{\xi^2+|\Delta|^2} \right) ,
\label{gapeq2}
\eeq
where we have taken $\Theta_\k=\Theta(|\xik|-\Lambda)$. The density of states
in the normal phase $N(\xi)=\frac{1}{V} \sum_\k \delta(\xi-\xi_\k)$ 
has been approximated by its value $N(0)$ at the Fermi level, and  
$\tilde\lambda=N(0)\lambda$ is a dimensionless interaction constant.
Eq.~(\ref{gapeq2}) can be solved exactly at $T=0$, 
\beq
|\Delta(T=0)| = |\Delta_0| \left(1 - \frac{2\Lambda}{|\Delta_0|}\right)^{1/2}
\Theta\left(\frac{|\Delta_0|}{2}-\Lambda\right),  
\label{DeltaT0} 
\eeq 
assuming $|\Delta|\ll \Lambda_0$. $|\Delta_0|=|\Delta|_{\Lambda=0}=2\Lambda_0
e^{1/\tilde\lambda}$ is the zero-temperature BCS gap.  

When $\Delta(T,\Lambda)=0$, the singlet response function $\chi_{\rm
s}(x,x,y,y)$ defined in Sec.~\ref{subsec:RFN} is given by the Bethe-Salpeter
equation (\ref{chi}). In Fourier space,  
\beq
\chi_{\rm s}(q) = \frac{2 \Pi^{-+}(q)}{1+\lambda \Pi^{-+}(q)}
\eeq 
($q=(\q,i\wnu)$ with $\wnu$ a bosonic Matsubara frequency),
where $\Pi^{-+}(q)$ is the Fourier transform of the pp
propagator
\beq
\Pi^{-+}(x-y) = [C(x,y)]^2 .
\eeq 
The condition for $\chi_{\rm s}(q=0)$ to diverge, $1+\lambda \Pi^{-+}(q=0)=0$,
is equivalent to the linearized gap equation [Eq.~(\ref{gapeq2}) with
$\Delta\to 0^+$]. 

\begin{figure}
\includegraphics[bb=70 490 270 610,width=6cm]{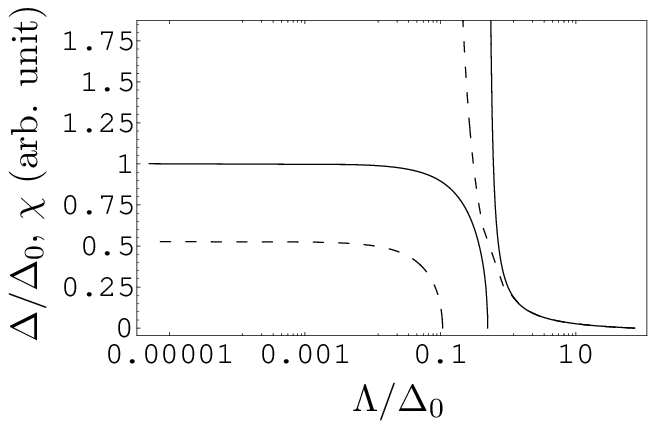}
\caption{Order parameter $\Delta$ and singlet response
  function $\chi_{\rm s}$ (for $\Delta(T,\Lambda)=0$) versus $\Lambda$ for
  $T=0$ (solid lines) and $T=0.9 T_c$ (dashed lines). }
\label{fig:op_bcs}
\end{figure}

Fig.~\ref{fig:op_bcs} shows the superconducting order parameter $\Delta$ and
the singlet response function $\chi_{\rm s}(q=0)$ (in the regime where
$\Delta(T,\Lambda)=0$) 
at $T=0$ and $T=0.9T_c$ where $T_c$ is the superconducting transition
temperature. $\chi_{\rm s}$ diverges at the threshold value $\Lambda_c(T)$ of
the cutoff below which the gap $\Delta$ becomes finite and reaches the BCS
value for $\Lambda=0$. The transition temperature is determined by
$\Lambda_c(T_c)=0$. As the 2PI RG equations determine $\Delta$ and $\Phid$,
they are not plagued with divergencies and can be continued down to
$\Lambda=0$ for any temperature. Broken symmetry is signaled by a finite value
of the anomalous self-energy $\Delta$ below $\Lambda_c(T)$. 
The divergence of the singlet
response function $\chi_{\rm s}$ is seen only when the 2PI vertex $\Phid$ is
fed into the Bethe-Salpeter equation relating $\chi_{\rm s}$ to
$\Phi^{(2)}_{\rm pp}$. 

Note that we have picked up the nonzero solution of (\ref{gapeq11}) by
hand. Alternatively -- and this is how one should proceed in more complicated
situations -- one can directly solve the flow equation
(\ref{dotdelta}),\cite{note8} 
\begin{multline}
\frac{\dot\Delta}{\tilde\lambda} = \frac{\Delta}{\sqrt{\Lambda^2+\Delta^2}}
\tanh \left( \frac{\beta}{2} \sqrt{\Lambda^2+\Delta^2} \right) \\
- \dot\Delta \int_\Lambda^{\Lambda_0} \frac{d\xi}{E} \tanh \left(
\frac{\beta}{2} E \right) \\
+ \dot\Delta \Delta^2  \int_\Lambda^{\Lambda_0} \frac{d\xi}{E^2} \left[
  \frac{1}{E} 
  \tanh \left( \frac{\beta}{2} E \right) - \frac{\beta}{2
    \cosh^2\left(\beta E/2 \right)} \right] 
\label{gapeq111}
\end{multline}
($E=\sqrt{\xi^2+\Delta^2}$ and we assume the gap $\Delta$ to be real),
together with a symmetry-breaking initial condition
$\Delta(\Lambda_0)=\eps$. For $\eps/\Delta(\Lambda=0)\lesssim 10^{-5}$, the
solution of (\ref{gapeq111}) cannot be distinguished from that of
(\ref{gapeq2}) when plotted on the same graph. A larger value of $\eps$ leads
to a smearing of the singularity at $\Lambda_c(T)$.

\subsection{Thermodynamic potential}

We could determine the thermodynamic potential $\Omega=\beta^{-1}\Gamma$
directly from Eq.~(\ref{phi_def}) using the expression (\ref{phi_bcs}) of
$\Phi_{\rm BCS}$. To illustrate how the 2PI RG scheme works, we shall instead
use the flow equation (\ref{Omega_rg}) and the Ginzburg-Landau expansion
(\ref{GL}) of the thermodynamic potential near a phase transition. 

\subsubsection{Ground state condensation energy}
\label{subsubsec_gsce} 

The ground state condensation energy
$\Delta\Omega=\Omega - \Omega_0$ satisfies the RG equation [see
  Eq.~(\ref{Omega_rg})] 
\bleq
\Delta\dot\Omega &=& - \frac{1}{2\beta} \Tr[\dot C^{-1}(G-C)] \nonumber \\
&=& - \frac{2}{\beta} \sum_k \dot C^{-1}_k[G_{\up\up}(k-,k+)-C(k)] \nonumber
\\ 
&=& - \frac{2}{\beta} \sum_k
\frac{|\Delta|^2\dot\Theta_\k\Theta_\k}{\wn^2+E^2_\k} \nonumber \\
&=& - |\Delta|^2 \sum_\k \frac{\dot\Theta_\k\Theta_\k}{E_\k} ,
\label{gse1}
\eleq
where the last line is obtained for $T=0$. A direct evaluation of this
equation by replacing the sum over $\k$ by an integral over $\xi_\k$ leads to
ambiguities because of factors $\Theta(0)$. We therefore rewrite
Eq.~(\ref{gse1}) as
\beq
\Delta\dot\Omega = - \bar\partial_\Lambda \sum_\k
\sqrt{\xi^2_\k+|\Delta|^2\Theta^2_\k} ,
\eeq
where $\bar\partial_\Lambda$ denotes a derivation with respect to $\Lambda$
at fixed $|\Delta|$. The calculation is then straightforward and yields
\bleq
\Delta\dot\Omega &=& -2VN(0)\Bigl[\Lambda-\sqrt{\Lambda^2+|\Delta|^2}\Bigr]
 \nonumber \\ 
&=& -2VN(0)(2\Lambda-|\Delta_0|)
 \Theta\biggl(\frac{|\Delta_0|}{2}-\Lambda\biggr), 
\eleq
where we have used Eq.~(\ref{DeltaT0}). Integrating this equation between
$\Lambda_0$ and $\Lambda$, we finally obtain
\beq
\Delta\Omega = - 2VN(0)\biggl(\frac{|\Delta_0|}{2}-\Lambda\biggr)^2
\Theta\biggl(\frac{|\Delta_0|}{2}-\Lambda\biggr) .
\eeq
For $\Lambda=0$, we recover the expression of the condensation energy
$\Delta\Omega=-\half VN(0)|\Delta_0^2|$ in a BCS superconductor. 

\subsubsection{Ginzburg-Landau expansion} 

In the vicinity of the superconducting transition, the thermodynamic potential
can be calculated using the general method discussed in Sec.~\ref{subsec:GL}. 
This approach relies on the 2PI RG
equations in the normal phase ($\Delta=0$) continued below the actual
transition temperature $T_c$ (i.e. with $\Sigma=\Sigma_N=0$ and $\Phid={\rm
const}$ in the BCS model). The anomalous self-energy $\Delta_{\gam_1\gam_2}$ is
defined by 
\bleq
\Delta_{\sig_1\sig_2}(x_1c_1,x_2c_2) &=& \delta(x_1-x_2)
\delta_{\sig_1,\bar\sig_2} \delta_{c_1,c_2} \nonumber \\ && \times 
\sig_1 (\delta_{c_1,+} \Delta - \delta_{c_1,-} \Delta^*) .
\label{Sig_ano}
\eleq
Using (\ref{Sig_ano}), one finds
\bleq
\Tr(G_N\Delta)^2 &=& -4 |\Delta|^2 \sum_k C(k)C(-k) \nonumber \\
               &=& -4 V\beta |\Delta|^2 N(0) \ln\left(
\frac{2\gamma\Lambda_0}{\pi T} \right) , \nonumber \\
\Tr(G_N\Delta)^4 &=& 4 |\Delta|^4 \sum_k [C(k)C(-k)]^2 \nonumber \\
               &=& V\beta |\Delta|^4 \frac{7\zeta(3)N(0)}{2\pi^2 T^2}
\nonumber \\ 
\frac{1}{8} \sum_{\gam_1,\gam_2} F^{(2)}_{\gam_1\gam_2} \Delta_{\gam_1}
\Delta_{\gam_2} &=& \frac{|\Delta|^2}{2} \int dx_1 dx_2
F^{(2)}_{\rm s}(x_1,x_1,x_2,x_2) \nonumber \\ 
&=& - V\beta \frac{|\Delta|^2}{\lambda} ,
\label{traces_bcs}
\eleq
where $\gamma\simeq 1.78$ is the exponential of the Euler constant and
$\zeta(z)$ the Riemann zeta function ($\zeta(3)\simeq 1.2$). 
To obtain the last line, we have used the relation (\ref{Phi_F}) between
$\Phi^{(2)}_{\rm s}$ and $F^{(2)}_{\rm s}$ and Eq.~(\ref{Phi2_bcs}). From
Eqs.~(\ref{GL},\ref{traces_bcs}), we recover the Ginzburg-Landau expansion
in a BCS superconductor,
\beq
\frac{\Delta\Omega}{V} = |\Delta|^2\biggl[ \frac{1}{|\lambda|} - N(0) \ln
  \left( \frac{2\gamma\Lambda_0}{\pi T} \right)\biggr]
+ \frac{7\zeta(3)N(0)}{16\pi^2 T^2} |\Delta|^4 .
\label{GL_bcs} 
\eeq

\subsection{Response function and collective modes in the superconducting
  phase} 

For $\Lambda<\Lambda_c(T)$, where $\Delta(T,\Lambda)\neq 0$, the singlet
superconducting response function is defined by  
\bleq
\chi^{c_1c_2}_{\rm s}(x_1,x_2) &=& \half \sum_{\sig_1,\sig_2} \sig_1\bar\sig_2
\nonumber \\ && \times 
W^{(2)}_{\sig_1\bar\sig_1\sig_2\bar\sig_2}(x_1,c_1,x_1,c_1,x_2,c_2,x_2,c_2) .
\nonumber \\ && 
\eleq
Using the Bethe-Salpeter equation (\ref{bethe}), we obtain 
\bleq
\chi^{-+}_{\rm s}(q) &=& 2 \Pi^{-+}(q) - \lambda 
\Pi^{-+}(q) \chi^{-+}_{\rm s}(q) \nonumber \\ && 
- \lambda \Pi^{--}(q) \chi^{++}_{\rm s}(q) , \nonumber \\ 
\chi^{++}_{\rm s}(q) &=& 2 \Pi^{++}(q) - \lambda 
\Pi^{+-}(q) \chi^{++}_{\rm s}(q)  \nonumber \\ &&
- \lambda \Pi^{++}(q) \chi^{-+}_{\rm s}(q) ,
\eleq 
where $\Pi^{c_1c_2}(q)$ is the Fourier transform of 
\bleq
\Pi^{-+}(x-y) &=& \Pi^{+-}(y-x) = G_{\up\up}(x,y) G_{\down\down}(x,y) , 
\nonumber \\ 
\Pi^{--}(x-y) &=& - G_{\down\up}(x-,y-) G_{\up\down}(x-,y-) ,
\nonumber \\
\Pi^{++}(x-y) &=& - G_{\up\down}(x+,y+) G_{\down\up}(x+,y+) .
\eleq
Collective modes are obtained from the poles of $\chi^{c_1c_2}(q)$, 
\beq
[1+\lambda \Pi^{-+}(q)][1+\lambda \Pi^{+-}(q)] 
-\lambda^2 \Pi^{++}(q)\Pi^{--}(q) = 0 .
\label{modes}
\eeq
Using 
\bleq
\Pi^{-+}(q=0) &=& \Pi^{+-}(q=0) \nonumber \\ 
&=& - \frac{1}{\lambda} - \frac{1}{\beta V}
\sum_k \frac{\Theta^2_\k |\Delta^2_\k|}{(\wn^2+\Ek^2)^2} , \nonumber \\ 
\Pi^{--}(q=0) &=& \Pi^{++}(q=0)^* \nonumber \\ &=&  \frac{1}{\beta V}
\sum_k \frac{\Delta^2_\k}{(\wn^2+\Ek^2)^2} ,
\eleq
one verifies that Eq.~(\ref{modes}) is satisfied for $\q=0$ and
$\wnu=0$. For any finite value of the gap $\Delta$ (i.e. for
$\Lambda<\Lambda_c$), we therefore obtain a gapless (Goldstone) mode
(Anderson-Bogoliubov mode\cite{Anderson58,Bogoliubov58}). 
Note that the Goldstone theorem is ensured by the fact
that the BCS theory is a $\Phi$-derivable approximation.

\section{One-dimensional systems}
\label{sec:LL}

In this section, we consider a 1D system with the action (for a review on 1D
systems, see Ref.~\onlinecite{Bourbonnais95})  
\bleq
S_0 &=& -\sum_{k,r,\sig} \psi^*_{r\sig}(k)
[i\omega_n-\xi_r(\kpara)] \psi_{r\sig}(k) , \nonumber \\ 
S_{\rm int} &=& \frac{1}{2\beta L} 
\sum_{k,k',q \atop r,r',\sig,\sig'} 
(g_2 \delta_{r,r'} + g_1 \delta_{r,\bar r'}) 
\nonumber \\ && \times 
\psi^*_{r\sig}(k+q) \psi^*_{\bar r\sig'}(k'-q) 
\psi_{\bar r'\sig'}(k') \psi_{r'\sig}(k) ,
\label{S1D}
\eleq 
where $L$ is the length of the system, $k=(\kpara,i\wn)$, and
$q=(q_\parallel,i\omega_\nu)$. $\kpara$ 
and $q_\parallel$ denote momenta, $\wn$ and $\w_\nu$ fermionic and bosonic 
Matsubara frequencies, respectively. The index $r$ distinguishes
between right ($r=+$) and left ($r=-$) moving
fermions. $\xi_r(\kpara) = \epsilon_r(\kpara)-\mu =
v_F(r\kpara-k_F)$ is the dispersion law, linearized around the two Fermi
points $\pm k_F$, $\mu$ being the chemical potential. The bandwidth is
$2\Lambda_0=2\,{\rm max} |\xi_r(\kpara)|$. $g_1$ and $g_2$ are the
backward and forward scattering amplitudes, respectively. We assume the
band filling to be incommensurate and neglect Umklapp processes. 

\begin{figure}
\includegraphics[width=8cm]{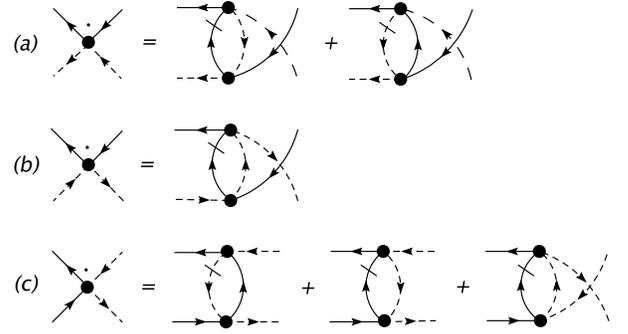}
\caption{One-loop RG equations for the 2PI vertex $\Phid$ in the Cooper (a),
  Peierls (b) and Landau (c) channels in a 1D
  system. Solid and dashed lines indicate right- and left-moving fermions,
  respectively. [In the pp loops appearing in (b) and (c), a
  sum over the left/right indices is implied.] }
\label{fig:rg1D}
\end{figure}

\subsection{One-loop RG equations}

The 2PI two-particle vertex $\Phi^{(2)r_1r_2r_3r_4}_{\rm
pp,\rm ph}(X_i)$ now carries left/right indices $r_i$. Momentum
conservation implies $\sum_i r_i=0$. We define
\bleq
\Phi^{\rm C}_\mu(x_i) = \Phi^{(2)+--+}_\mu(x_i) , && (\mu={\rm
  t,s}),  \nonumber \\ 
\Phi^{\rm P}_\mu(x_i) = \Phi^{(2)+--+}_\mu(x_i) , && (\mu={\rm
  ch,sp}), \nonumber \\ 
\Phi^{\rm L}_\mu(x_i) = \Phi^{(2)++--}_\mu(x_i) , && (\mu={\rm
  ch,sp}), 
\label{phiCPL1} 
\eleq 
where C, P, and L refer to the Cooper, Peierls, and Landau
channels according to the standard terminology used in the framework of the
g-ology model.\cite{Bourbonnais95} In Fourier space, we neglect the frequency
dependence and approximate  
\bleq
\Phi^{\rm C}_\mu(k_1,k_2,k_3,k_4) &\simeq& \Phi^{\rm C}_\mu
(k_F,-k_F,-k_F,k_F) \equiv \Phi^{\rm C}_\mu       , \nonumber \\ 
\Phi^{\rm P}_\mu(k_1,k_2,k_3,k_4) &\simeq& \Phi^{\rm P}_\mu
(k_F,-k_F,-k_F,k_F) \equiv \Phi^{\rm P}_\mu , \nonumber \\ 
\Phi^{\rm L}_\mu(k_1,k_2,k_3,k_4) &\simeq& \Phi^{\rm L}_\mu
(k_F,k_F,-k_F,-k_F) \equiv \Phi^{\rm L}_\mu . \nonumber \\ && 
\label{phiCPL2}
\eleq 
By scaling arguments, one can show that the dependence of the 1PI vertex
$\gamma^{(4)}(k_i)$ on $\w_{n_i}$ and $|k_{i\para}|-k_F$ is irrelevant in the
RG sense. The 
validity of Eq.~(\ref{phiCPL2}), which assumes that the 2PI vertex $\Phid$
shares the same property, will be discussed in Sec.~\ref{subsec:LL_disc}. 
One-loop flow equations for $\Phi^{\rm C}$, $\Phi^{\rm P}$
and $\Phi^{\rm L}$ are deduced from Eqs.~(\ref{rg_eq4}) by including the
$r$ index, i.e. $x_i\to (x_i,r_i)$. Retaining only leading
logarithmic divergent loops (Parquet approximation),\cite{Bourbonnais95} the
contribution to the self-energy vanishes and the RG equations for the
two-particle vertex read (see Fig.~\ref{fig:rg1D}) 
\bleq
{\dot\Phi}^{\rm C}_{\rm t} &=& - B_{\rm ph} \bigl(\PhiCt\PhiLch + 2
\PhiCt\PhiLsp - \PhiCs\PhiLsp \bigr) , \nonumber \\ 
{\dot\Phi}^{\rm C}_{\rm s} &=& - B_{\rm ph} \bigl(\PhiCs\PhiLch - 3
\PhiCt\PhiLsp \bigr) , \nonumber \\  
{\dot\Phi}^{\rm P}_{\rm ch} &=&  -B_{\rm pp} \bigl(\PhiPch\PhiLch + 3
\PhiPsp\PhiLsp \bigr) , \nonumber \\  
{\dot\Phi}^{\rm P}_{\rm sp} &=&  -B_{\rm pp} \bigl(\PhiPch\PhiLsp +
\PhiPsp\PhiLch - 2 \PhiPsp\PhiLsp \bigr) , \nonumber \\
{\dot\Phi}^{\rm L}_{\rm ch} &=&  -\half B_{\rm ph} \biggl( \frac{3}{4}
{\PhiCt}^2 + \quarter {\PhiCs}^2 + {\PhiLch}^2 + 3 {\PhiLsp}^2 \biggr) 
\nonumber \\ && 
-\half B_{\rm pp} \Bigl( {\PhiLch}^2 + 3 {\PhiLsp}^2 + {\PhiPch}^2 +3
{\PhiPsp}^2 \Bigr) , 
\nonumber \\
{\dot\Phi}^{\rm L}_{\rm sp} &=&  - B_{\rm ph} \biggl( \quarter
{\PhiCt}^2 - \quarter \PhiCt\PhiCs + \PhiLch\PhiLsp + {\PhiLsp}^2 \biggr) 
\nonumber \\ && 
- B_{\rm pp} \Bigl( \PhiLch\PhiLsp -  {\PhiLsp}^2 + 
\PhiPch\PhiPsp -   {\PhiPsp}^2 \Bigr) , 
\label{flow1D}
\eleq 
where
\bleq
B_{\rm ph} &=& \frac{1}{\beta L} \sum_{k} [ G_+(k)
  \dot G_-(k-Q) + (G \leftrightarrow \dot G) ] , 
\nonumber \\ 
B_{\rm pp} &=& \frac{1}{\beta L} \sum_{k} [ G_+(k)
  \dot G_-(-k) + (G \leftrightarrow \dot G) ] 
\eleq
come from the ph and pp loops, and $Q=(2k_F,0)$. The dot denotes a
derivation with respect to $l=\ln(\Lambda_0/\Lambda)$. To
evaluate $B_{\rm ph}$ and $B_{\rm pp}$, we use a sharp infrared cutoff,  
\beq
G_r(k) = C_r(k) = - \frac{\Theta(|\xi_r(\kpara)|
-\Lambda)}{i\wn-\xi_r(\kpara)} .
\eeq
This gives
\beq
B_{\rm pp} = - B_{\rm ph} = \frac{1}{2\pi v_F} \tanh\left(\beta
\frac{\Lambda}{2} \right) .
\eeq 
The initial values of the 2PI vertex $\Phid$ is the bare vertex defined by
Eq.~(\ref{S1D}), 
\bleq
\PhiCt |_{\Lambda_0} = -g_1+g_2 , &&  
\PhiCs |_{\Lambda_0} =  g_1+g_2 , \nonumber \\ 
\PhiPch |_{\Lambda_0} = g_1-\frac{g_2}{2} , &&  
\PhiPsp |_{\Lambda_0} =  -\frac{g_2}{2} , \nonumber \\ 
\PhiLch |_{\Lambda_0} = -\frac{g_1}{2}+g_2 , &&  
\PhiLsp |_{\Lambda_0} =  -\frac{g_1}{2} . 
\label{init1D} 
\eleq

\subsection{Response functions and 1PI vertices} 

The response functions in the Cooper and Peierls channels are defined by
\bleq
\chi^{\rm C}_{\rm t,s} &=& \frac{1}{\beta L} \sum_{k,k'} 
\chi_{\rm t,s}^{+--+}(k,-k,-k',k') , \nonumber \\ 
\chi^{\rm P}_{\rm ch,sp} &=& \frac{1}{\beta L} \sum_{k,k'} 
\chi_{\rm ch,sp}^{+--+}(k,k-Q,k'-Q,k') .
\eleq
From the Bethe-Salpeter equations satisfied by $\chi_{\rm t,s}^{+--+}$ and
$\chi_{\rm ch,sp}^{+--+}$ (Eqs.~(\ref{chi}) with the $r_i$ index
included), we obtain 
\bleq
\chi^{\rm C}_\mu &=& \frac{\Pi_{\rm pp}}{1+\Pi_{\rm pp} \Phi^{\rm C}_\mu} , 
\nonumber \\
\chi^{\rm P}_\mu &=& \frac{-2\Pi_{\rm ph}}{1-2\Pi_{\rm ph} \Phi^{\rm P}_\mu} ,
\label{chi_1D}
\eleq
where 
\bleq
\Pi_{\rm pp} &=& \frac{1}{\beta L} \sum_k G_+(k) G_-(-k) \nonumber \\ 
             &=& \frac{1}{2\pi v_F} \int_\Lambda^{\Lambda_0}
\frac{d\xi}{\xi} \tanh \left(\beta\frac{\xi}{2} \right) , \nonumber \\ 
\Pi_{\rm ph} &=& \frac{1}{\beta L} \sum_k G_+(k) G_-(k-Q) \nonumber \\
             &=& - \Pi_{\rm pp} .
\label{Pipp1D} 
\eleq
Similarly, for the 1PI two-particle vertex $\gamma^{(4)}$
[Eq.~(\ref{bethe_gamma4})], we find  
\bleq
\gamma^{\rm C}_\mu &=& \frac{\Phi^{\rm C}_\mu}{1+\Pi_{\rm pp}
  \Phi^{\rm C}_\mu} , \nonumber \\ 
\gamma^{\rm P}_\mu &=& \frac{\Phi^{\rm P}_\mu}{1-2\Pi_{\rm ph}
  \Phi^{\rm P}_\mu} ,\nonumber \\ 
\gamma^{\rm L}_\mu &=& \Phi^{\rm L}_\mu . 
\label{gamma4_1D}
\eleq
The equality between $\gamma^{\rm L}_\mu$ and $\Phi^{\rm L}_\mu$ is due to the
absence of logarithmic divergent loops in the Landau channel and
holds at the Parquet level. 

In Sec.~\ref{sec:formalism}, we have shown very generally that the RG
equation for $\gamma^{(4)}$ derived within the 1PI scheme follows from the RG
equation satisfied by the 2PI vertex $\Phid$ and the Bethe-Salpeter equation
relating $\gamma^{(4)}$ and $\Phid$. In appendix \ref{app:LL}, as a means to
check the validity of Eqs.~(\ref{flow1D}), we recover the RG equation
satisfied by $\gamma^{(4)}$ directly from Eqs.~(\ref{flow1D},\ref{gamma4_1D}).

\begin{figure}[tp]
\includegraphics[width=7.cm,bb=95 230 315 792]{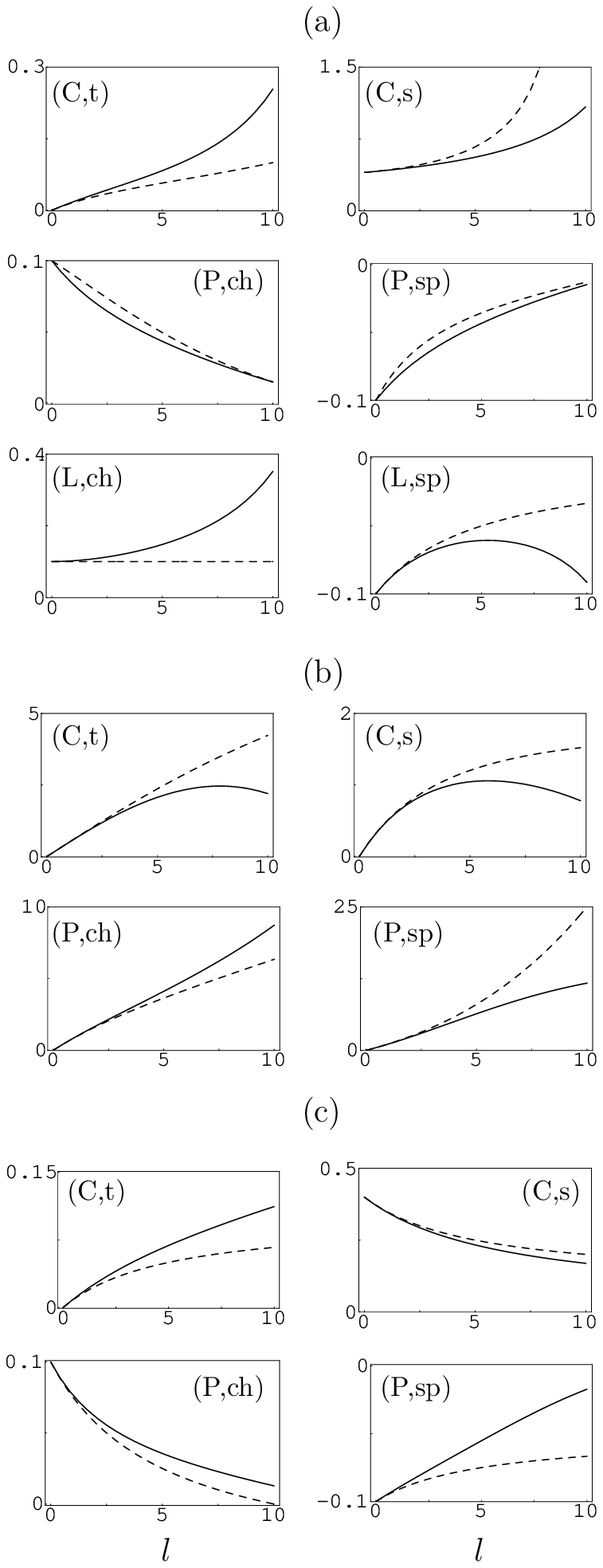}
\caption{2PI vertices $\Phid$ (a), response functions $\chi$ (b) and 1PI
  vertices $\gamma^{(4)}$ (c) versus $l=\ln(\Lambda_0/\Lambda)$ for $\tilde
  g_1=\tilde g_2=0.2$ and $T=0$. Units are chosen such that $\pi v_F=1$. 
  The correlation channel is indicated in the
  upper left or right corner of each graph, following the notation of the text:
  (C,t)=(Cooper,triplet), etc.   
  The solid (dashed) lines show the results obtained from the 2PI (1PI)
  RG scheme. In the 1PI RG scheme, the 2PI vertex $\Phid$ (dashed lines in
  panel (a)) is deduced from $\gamma^{(4)}$ using the relations
  (\ref{gamma4_1D}). }
\label{fig:LL1}
\end{figure}
\begin{figure}[tp]
\includegraphics[width=7.cm,bb=95 230 315 720]{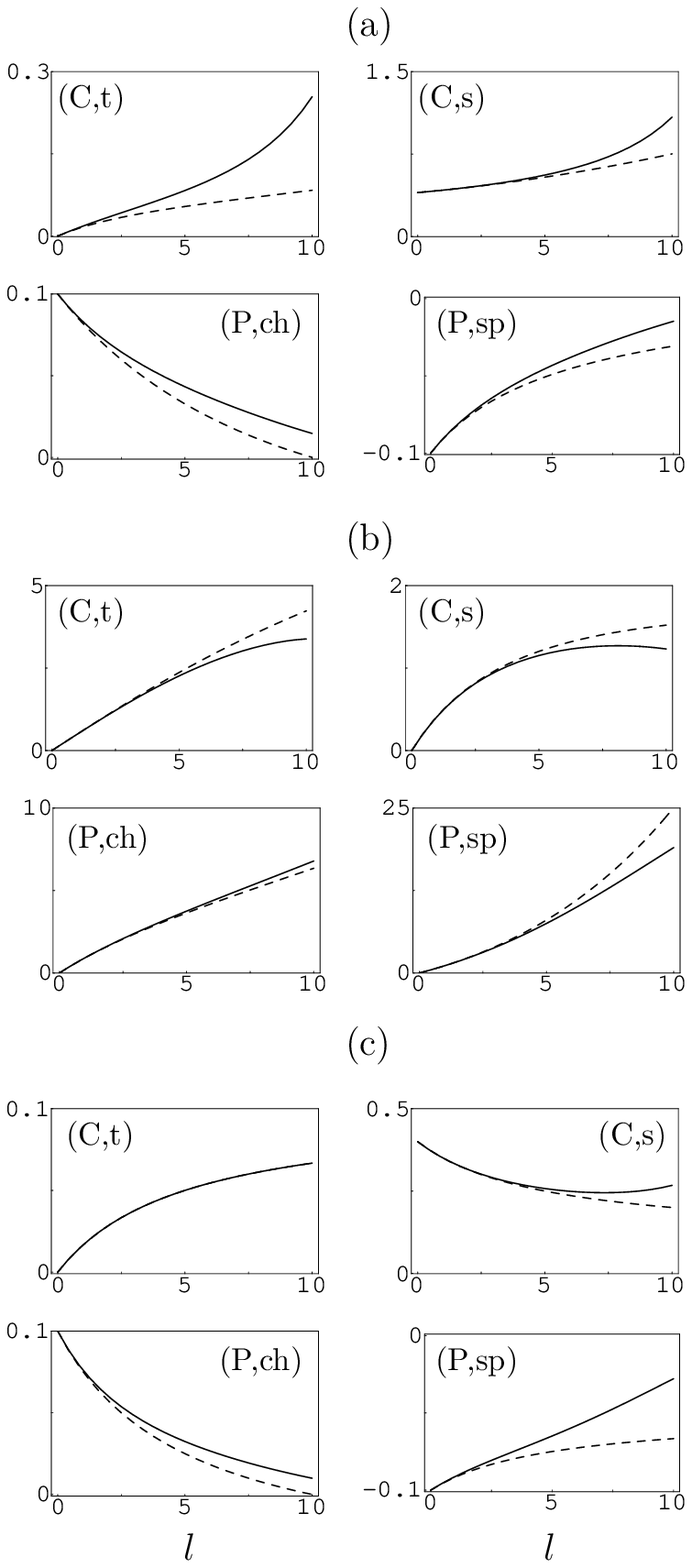}
\caption{Same as Fig.~\ref{fig:LL1}, but using Eqs.~(\ref{improve}) instead of
  Eqs.~(\ref{chi_1D},\ref{gamma4_1D}). }
\label{fig:LL2} 
\end{figure}

\subsection{Discussion}
\label{subsec:LL_disc}

Figure \ref{fig:LL1} shows $\Phid$, $\chi$ and $\gamma^{(4)}$ versus
$l=\ln(\Lambda_0/\Lambda)$ at zero temperature for $\tilde g_1=\tilde
g_2=0.2$. $\tilde g_1=g_1/\pi v_F$ and $\tilde g_2=g_2/\pi 
v_F$ are dimensionless coupling constants. Solid lines show the results
obtained within the 2PI scheme by solving the flow equations (\ref{flow1D})
and using Eqs.~(\ref{chi_1D},\ref{gamma4_1D}). Dashed lines correspond to
results obtained within the 1PI scheme, where $\gamma^{(4)}$ and $\chi$ are
directly obtained from RG equations, the 2PI vertex $\Phid$ being then deduced
from the relations (\ref{gamma4_1D}). The agreement between the two schemes
is excellent at high energies (small $l$), but deteriorates at lower
energies where, in at least one correlation channel, the 2PI vertex $\Phid$
becomes of order one and eventually diverges. This deficiency is not
important in the 1PI scheme, as $\Phid$ is usually not considered in this
scheme, but it shows that the one-loop approximation breaks down in the 2PI
scheme at low energy. 

The divergence of the 2PI vertex $\Phid$ in the 1PI scheme, which correctly
predicts all other physical quantities, suggests 
that the relations (\ref{gamma4_1D}) between 1PI and 2PI vertices may not be
quite correct. Inverting Eqs.~(\ref{gamma4_1D}) and considering the zero
temperature limit, we obtain $\tilde\Phi^{\rm C}_\mu(l)= \tilde\gamma^{\rm
  C}_\mu(l)/(1-l\tilde\gamma^{\rm C}_\mu/2)$ in the Cooper channel and $\tilde
\Phi^{\rm P}_\mu(l)=\tilde\gamma^{\rm P}_\mu(l)/(1-l\tilde\gamma^{\rm P}_\mu)$
in the Peierls channel. $\tilde\Phi=\Phi/\pi v_F$ and $\tilde\gamma=
\gamma/\pi v_F$ are dimensionless vertices. 
We conclude that the 2PI vertex $\Phid$ will diverge at a finite energy
scale $\Lambda_c=\Lambda_0 e^{-l_c}$ whenever $\gamma^{(4)}$ is positive and
finite in the low-energy limit ($l\to\infty$). For repulsive interactions,
$\gamma^{\rm C}_{\rm s}(l)\to 
g_2-g_1/2$ satisfies this condition when $g_2>g_1/2$, which leads to the
divergence of the 2PI vertex $\Phi^{\rm C}_{\rm s}$ (see
Fig.~\ref{fig:LL1}). The unphysical divergence of the 2PI vertex
$\Phid$ comes from the assumption that the 2PI vertex is momentum 
independent [Eq.~(\ref{phiCPL2})], which results in an artificial decoupling
of energy scales between 
the 2PI vertex and the reducible pp (or ph)
propagator $\Pi_{\rm pp,ph}$. The latter involves all energy scales between
$\Lambda$ and $\Lambda_0$ whereas the momentum-independent $\Phid$ is an 
effective 2PI vertex at the energy scale $\Lambda$. The Bethe-Salpeter
equations (\ref{chi_1D}) and (\ref{gamma4_1D}) are therefore expected to be
less and less reliable as the infrared cutoff $\Lambda$ decreases. 

These difficulties can be partially overcome by modifying the Bethe-Salpeter
equation relating $\Phid$ to $\gamma^{(4)}$ and $\chi$,
\bleq
\tilde\gamma^{\rm C}_\mu(\Lambda) &=& \tilde\Phi^{\rm C}_\mu(\Lambda) -
\frac{1}{2} 
\int_\Lambda^{\Lambda_0} 
\frac{d\xi}{\xi} \tanh\left(\beta\frac{\xi}{2}\right) \tilde\Phi^{\rm
  C}_\mu(\Lambda) 
\tilde\gamma^{\rm C}_\mu(\Lambda) \nonumber \\ 
&\to& \tilde\Phi^{\rm C}_\mu(\Lambda) - \frac{1}{2}
\int_\Lambda^{\Lambda_0} 
\frac{d\xi}{\xi} \tanh\left(\beta\frac{\xi}{2}\right) \tilde\Phi^{\rm
  C}_\mu(\xi) 
\tilde\gamma^{\rm C}_\mu(\Lambda) , \nonumber \\
\chi^{\rm C}_\mu(\Lambda) &=&  \Pi_{\rm pp}(\Lambda) - \frac{1}{2}
\int_\Lambda^{\Lambda_0} 
\frac{d\xi}{\xi} \tanh\left(\beta\frac{\xi}{2}\right) \tilde\Phi^{\rm
  C}_\mu(\Lambda) 
\chi^{\rm C}_\mu(\Lambda) \nonumber \\ 
&\to&  \Pi_{\rm pp}(\Lambda) - \frac{1}{2} \int_\Lambda^{\Lambda_0}
\frac{d\xi}{\xi} \tanh\left(\beta\frac{\xi}{2}\right) \tilde\Phi^{\rm
  C}_\mu(\xi) \chi^{\rm C}_\mu(\Lambda) , \nonumber \\ && 
\eleq
and similar equations in the ph channel. The 2PI vertex $\Phid$ is now taken
at the same energy scale than the pp or ph propagator. We thus obtain
\bleq 
\gamma^{\rm C}_\mu = \frac{\Phi^{\rm C}_\mu}{1+\frac{\tilde\Psi^{\rm
      C}_\mu}{2}} , &&  
\chi^{\rm C}_\mu = \frac{\Pi_{\rm pp}}{1+\frac{\tilde\Psi^{\rm C}_\mu}{2}} ,
\nonumber \\ 
\gamma^{\rm P}_\mu = \frac{\Phi^{\rm P}_\mu}{1+\tilde\Psi^{\rm P}_\mu} , &&
\chi^{\rm P}_\mu = \frac{-2\Pi_{\rm ph}}{1+\tilde\Psi^{\rm P}_\mu} ,
\label{improve}
\eleq
where 
\beq
\Psi^{\rm C,P}_\mu = \int_\Lambda^{\Lambda_0} \frac{d\xi}{\xi} \tanh \left(
\beta \frac{\xi}{2}\right) \Phi^{\rm C,P}_\mu\Bigl|_{\Lambda=\xi} 
\eeq 
satisfies the flow equation 
\beq
\dot \Psi^{\rm C,P}_\mu = \tanh \left(\beta \frac{\Lambda}{2}\right) \Phi^{\rm
  C,P}_\mu .
\label{psi_flow}
\eeq 
Fig.~\ref{fig:LL2} shows $\Phid$, $\chi$ and $\gamma^{(4)}$ obtained from
Eqs.~(\ref{improve}) instead of Eqs.~(\ref{chi_1D},\ref{gamma4_1D}). The
parameters are the same as in Fig.~\ref{fig:LL1}. We see that the agreement
between the 1PI and 2PI schemes, in particular for the susceptibilities, is
significantly better. It is instructive to consider the case $g_1=0$ where 
the 1PI vertices $\gamma^{\rm C}_\mu$ and $\gamma^{\rm P}_\mu$ are
fixed points of the flow equations at any order in a loop expansion. From
Eqs.~(\ref{improve},\ref{psi_flow}), one then obtains $\Phi^{\rm C}_\mu =
\Phi^{\rm C}_\mu(l=0) \exp(\half \tilde\gamma^{\rm C}_\mu l)$ and $\Phi^{\rm
  P}_\mu = \Phi^{\rm P}_\mu(l=0) \exp(\tilde\gamma^{\rm P}_\mu l)$ at zero
temperature. Depending
on the sign of $\gamma_\mu$, the 2PI vertex will either vanish or diverge
exponentially in the limit $l\to\infty$. The unphysical divergence of the 2PI
vertex obtained earlier at a finite energy scale is now replaced by an
exponential divergence at zero energy ($\Lambda=\Lambda_0e^{-l}\to 0$). This
divergence follows from the coexistence of logarithmically divergent
(reducible) 
pp and ph loops and weak 1PI vertices, which is made
possible in 1D by the strong interferences between various correlation
channels. The 2PI flow equations give the correct qualitative behavior of
$\Phid$, but the one-loop approximation clearly breaks down at low energy when
$\Phid$ becomes of order one.

It should be noticed that if, 
within the 1PI scheme, the susceptibilities were calculated from 
\bleq
\chi^{\rm C}_\mu &=& \Pi_{\rm pp} - \Pi_{\rm pp} \gamma^{\rm C}_\mu
\Pi_{\rm pp} , \nonumber \\ 
\chi^{\rm P}_\mu &=& -2 \Pi_{\rm ph} - 4 \Pi_{\rm ph} \gamma^{\rm P}_\mu
\Pi_{\rm ph} , 
\label{chi_1Da}
\eleq
with momentum independent vertices $\gamma^{(4)}$ obtained from RG equations, 
similar difficulties would arise and 
the correct asymptotic low-energy behavior would not be reached. The correct
result is obtained by also deriving RG equations for the susceptibilities; in
this way irreducible and reducible pp and ph loops are
considered on equal footing. These RG equations are obtained by introducing
bosonic external sources that couple to order parameter fields, in addition to
the fermionic sources that are used to obtain the generating functional of 1PI
vertices by a Legendre transformation.\cite{Salmhofer01,Honerkamp01}   

To quantify the failure of the 2PI scheme, within the one-loop approximation,
to access the low-energy limit of 1D systems, let us consider the 
case of the quasi-1D organic conductors of the 
Bechgaard salt family. In these systems, the bandwidth $2\Lambda_0\simeq
4t_\para \sim 12000$ K is much larger than the kinetic interchain coupling
$t_\perp \sim 300$ K. For $\tilde g_1=\tilde g_2=0.2$, the 2PI flow equations
break down for $l\sim 5$ (Fig.~\ref{fig:LL2}), 
which corresponds to an energy or temperature scale
$\Lambda_0 e^{-l}\sim 40$ K. For $\tilde g_1=\tilde g_2=0.4$, we find that
the flow equations remain valid down to $\sim\Lambda_0 e^{-3}\sim 300$ K. 
In both cases, these energy scales are of the same order of magnitude or
smaller  than the temperature $T_x\sim t_\perp\sim 300$ K at which a crossover
to a 2D regime takes place. Preliminary calculations in quasi-1D systems
indicate that the 
interaction strength studied in Refs.~\onlinecite{Nickel05a,Nickel05b}, namely
$\tilde g_2=2\tilde g_1=0.64$ is accessible within the 2PI
scheme.\cite{Dupuis05} We thus conclude that the 2PI scheme can be used for
realistic quasi-1D systems like the Bechgaard salts. 

\section{Summary and conclusion} 

We have discussed the implementation of a Wilsonian momentum-shell RG approach
within the 2PI formalism introduced in the 60s by Luttinger, Ward, Baym and
others.\cite{Luttinger60,Baym61,Baym62,Dedominicis64}
The 2PI RG scheme yields an infinite hierarchy of flow equations satisfied by
the 2PI vertices $\Phi^{(n)}$. 
The susceptibilities are obtained from the Bethe-Salpeter equation that
relates them to the 2PI two-particle vertex $\Phid$. In the normal phase, one
has schematically 

\begin{picture}(80,85)
\setlength{\unitlength}{0.1cm}
\put(15,23){$\chi_{\rm pp} = \chi^{(0)}_{\rm pp} - \half \Pi_{\rm pp} \Phipp
  \chi_{\rm pp}$ }
\put(15,3){$\chi_{\rm ph} = \chi^{(0)}_{\rm ph} + 2 \Pi_{\rm ph} \Phiph
  \chi_{\rm ph}$ }
\put(46,16){\vector(0,1){4}}
\put(46,12){\vector(0,-1){4}}
\put(33,13){(RG equations)} 
\end{picture}

\noindent
(similar equations could be written for the 1PI vertices $\gamma^{(4)}_{\rm
pp}$ and $\gamma^{(4)}_{\rm ph}$) where $\chi^{(0)}_{\rm pp,ph}$ are the
susceptibilities without vertex correction and $\Pi_{\rm pp/ph}$ the pp or
ph pair propagator. The arrows indicate the coupling between the
pp and ph channels which is taken care of by the RG
equations satisfied by the 2PI two-particle vertices $\Phipp$ and $\Phiph$. The
infinite hierarchy of RG flow equations satisfied by the 2PI vertices
  $\Phi^{(n)}$ should be truncated at some order in a loop
expansion. The simplest non trivial truncation, the one-loop
approximation, was discussed in detail in Sec.~\ref{sec:formalism}.
As any approximation of the 2PI vertex $\Phid$, it leads to a violation of the
crossing symmetries of the two-particle Green function $W^{(2)}$ and the 1PI
vertex $\gamma^{(4)}$. Besides, the one-loop approximation is not a
$\Phi$-derivable approximation, and a detailed study of conservation laws and
Ward identities remains to be done.  

We have shown in Sec.~\ref{sec:LL} that 1D conductors are characterized by
an exponentially divergent 2PI vertex in the zero-energy limit, a consequence
of the strong interferences between correlation channels in 1D. 
This leads to a breakdown of the one-loop approximation, which
is therefore unable to access the asymptotic low-energy behavior of
Luttinger liquids. Nevertheless, we have argued that the 2PI scheme can be used
in quasi-1D systems like the organic conductors of the Bechgaard family where
a dimensional crossover always drives the system towards a 2D or 3D behavior
at low energy. 

On the other hand, the 2PI RG scheme leads to a particularly simple description
of single-channel (i.e. mean-field) theories. In mean-field theories, all the
$\Lambda$ dependence comes from the two-particle-reducible part of the
susceptibilities (and the 1PI vertex $\gamma^{(4)}$), while the 2PI vertex
$\Phid$ is invariant under the RG transformation: $\dot\Phi^{(2)}=0$.  The
one-loop flow equations then reduce to a single equation that determines the
(anomalous) self-energy and reproduces the usual mean-field gap equation
(Sec.~\ref{sec:BCS}). 

The possibility to continue the RG flow into broken-symmetry phases is an
essential feature of the 2PI RG scheme and is due to the fact that the 2PI
two-particle vertex $\Phid$, contrary to its 1PI counterpart, is not singular
at the phase transition. This property, which is obvious in a mean-field
theory, deserves some discussion in more general cases. 
The interchannel coupling, even weak, is likely to induce singularities in 
the 2PI vertex $\Phid$, in particular in the vicinity of a phase
transition. However, these singularities can be controlled by a proper
parameterization of $\Phid$. To see this, let us consider a quasi-1D
conductor near a 
spin-density-wave (SDW) instability. The spin susceptibility $\chi_{\rm
sp}(2k_F,\pi)$ diverges at the SDW transition but, as in single-channel
(mean-field) theories, the irreducible vertex $\Phi^{\rm P}_{\rm sp}$ should
remain finite (see Sec.~\ref{sec:LL} for the definition of $\Phi^{\rm P}$ and
$\Phi^{\rm C}$ in a (quasi-)1D system). In the vicinity of the phase
transition, nearly divergent spin fluctuations strongly affect the 2PI vertex
$\Phi^{\rm C}$ in the Cooper channel. For the purpose of our discussion, let
us assume that $\Phi^{\rm C}$ is proportional to the spin susceptibility,
\beq
\Phi^{\rm C}_{\mu={\rm t,s}}(k_\perp,q_\perp-k_\perp,
-k'_\perp,k'_\perp-q_\perp) \propto \chi_{\rm sp}(2k_F,k_\perp+k'_\perp) ,
\label{conclu1}
\eeq
where we retain the $k_\perp$ dependence of the vertex.\cite{Nickel05b} 
The combination of $k_\perp$ arguments in Eq.~(\ref{conclu1}) is the one that
appears in the Bethe-Salpeter equation determining the superconducting
susceptibility. Expanding the even function $\chi_{\rm sp}(2k_F,q_\perp)$ in
Fourier series, we obtain 
\bleq
\lefteqn{\Phi^{\rm C}_\mu (k_\perp,q_\perp-k_\perp,
-k'_\perp,k'_\perp-q_\perp)} &&  \nonumber \\ && 
= \sum_{n=0}^\infty a_\mu^{(n)}
\cos[n(k_\perp+k_\perp')] \nonumber \\ && 
= \sum_{n=0}^\infty  a_\mu^{(n)} [ \cos(nk_\perp)\cos(nk'_\perp)-
  \sin(nk_\perp)\sin(nk'_\perp) ] , \nonumber \\ && 
\eleq
where $a_\mu^{(n+1)}a_\mu^{(n)}<0$ and $|a_\mu^{(n)}|$ is a decreasing
function of $n$. The condition $|a_\mu^{(n)}|=|a_\mu^{(0)}|$,
i.e. $a_\mu^{(n)}=(-1)^n a_\mu^{(0)}$, would lead to a
diverging susceptibility $\chi_{\rm sp}(2k_F,q_\perp)\propto \delta(q_\perp
-\pi)$, while $|a_\mu^{(n+1)}|<|a_\mu^{(n)}|$ gives a
broadened peak at $q_\perp=\pi$. The proximity of the SDW transition
manifests itself by a larger and larger number of $a_\mu^{(n)}$ coefficients
with a significant amplitude. All these coefficients should however remain
bounded, with $1\pm(a_\mu^{(n)}/2)\Pi_{\rm pp}>0$, for the system to be stable
against a superconducting instability ($\Pi_{\rm pp}$ is the pp propagator
defined in (\ref{Pipp1D})). By parameterizing the 2PI vertex $\Phi^{\rm C}$ by
means of the $a_\mu^{(n)}$ coefficients, we avoid any complication due to the
diverging spin susceptibility. In practice, only a finite number of
coefficients need to be retained. The coefficients with a large value of $n$,
which correspond to pairing between fermions $n$ chain apart, do not play an
important role; they cannot drive a pairing instability -- an instability with
a small value of $n$ ($n=0,1,2,\cdots$) will always occur first\cite{note9}
-- and their influence on the ph channel is expected to be negligible. This
approximation will lead to a non essential smearing of the singularity of
$\Phi^{\rm C}_\mu$ at $k_\perp+k'_\perp=\pi$. The argument given here for
quasi-1D systems 
can be made more general. One can expand the 2PI vertex on the eigenmodes of
the Bethe-Salpeter equation, retaining only a finite number of
eigenmodes. Each coefficient in this expansion is bounded by a 
critical value at which a phase transition occurs. 

In Sec.~\ref{subsec:GL}, we have proposed a method to derive the
Ginzburg-Landau expansion of the thermodynamic potential in the vicinity of
the phase transition on the basis of the flow equations in the normal phase.
This is the simplest way to access phases with
long-range order as it does not require to solve the full RG equations in the
presence of symmetry breaking. The knowledge of the Ginzburg-Landau expansion
of the thermodynamic potential allows one to study the possibility of phase
coexistence below the transition temperature. This is a particularly important
issue in quasi-1D organic conductors where superconductivity and
antiferromagnetism,\cite{Vuletic02,Lee05} as well as spin- and charge-density
waves,\cite{Pouget96,Kagoshima99} coexist in some regions of the
pressure-temperature phase diagram. Recent 1PI RG calculations have indicated
that antiferromagnetism and superconductivity, as well as charge- and
spin-density-wave phases, lie nearby in the phase diagram of quasi-1D
conductors.\cite{Duprat01,Bourbonnais04,Nickel05a,Nickel05b} The 2PI RG scheme
would allow to 
determine whether they actually coexist in the low-temperature phase as
observed experimentally.\cite{Vuletic02,Lee05,Pouget96,Kagoshima99}

Finally, we note that the 2PI RG scheme enables a direct connection to the
phenomenological Landau
Fermi liquid theory when the metallic state remains stable down to low
temperature. It has been shown in Ref.~\onlinecite{Dupuis00} that the
functional $\Gamma[\bar G]$, or rather its variation $\delta\Gamma
=\Gamma[G]-\Gamma[\bar G]$, can be written as a functional
$\delta\Gamma[\delta n]$ of the Wigner distribution function $n=\lbrace
n_{\k\sig}(\r,\tau)\rbrace$ which is essentially determined by  
the Landau parameters. The latter
are given by the forward-scattering limit of the 2PI vertex
$\Phid$.  ($\delta n=n-\bar n$ denotes the
deviation from the equilibrium value $\bar n$.) $\delta\Gamma[\delta n]$
determines both static and dynamic 
properties of the Fermi liquid. In particular, it
yields the quantum Boltzmann equation satisfied by $n_{{\bf k}\sigma}({\bf
r},\tau)$. In the static case, the Wigner distribution $\lbrace n_{{\bf
    k}\sigma}({\bf r},\tau)\rbrace$ 
reduces to the quasi-particle distribution function $\lbrace n_{{\bf
k}\sigma}\rbrace$, and $\delta\Gamma[\delta n]=\delta E[\delta n]-\mu \delta N
- T \delta S[\delta n]$ where $\delta N[\delta n]$ and $\delta S[\delta n]$
are the quasi-particle number and entropy variations induced by a change
$\delta n$ in the quasi-particle distribution. The functional $\delta E[\delta
n]$, which gives the corresponding energy variation, is the basis of Landau's
phenomenological Fermi liquid theory. 

\begin{acknowledgments} 
I acknowledge discussions with M. Tissier on the 2PI formalism at an early
stage of this work. I am also grateful to C. Nickel and C. Bourbonnais for
numerous discussions on the RG approach, and to A. Katanin for a useful
comment on Ward identities.\cite{Katanin04}  
\end{acknowledgments} 

\appendix

\section{}
\label{app:LL}

In this appendix, we show that Eqs.~(\ref{flow1D}) correctly reproduce the
one-loop equations for the 1PI vertex $\gamma^{(4)}$ in Luttinger liquids. The
flow equation of $\gamma^{(4)}$ is given by 
Eq.~(\ref{gamma4_rg1}). To order $(\gamma^{(4)})^2$, it reduces to $\dot
\gamma^{(4)}=-\gamma^{(4)}\dot\Pi\gamma^{(4)}+{\dot\Phi}^{(2)}$ where
${\dot\Phi}^{(2)}$ should be evaluated to $\calO[(\gamma^{(4)})^2]$. This leads
to
\bleq
\dot\gamma^{\rm C}_{\rm t} &=& - B_{\rm pp} {\gamma^{\rm C}_{\rm t}}^2
+ {\dot\Phi}^{\rm C}_{\rm t}  \nonumber \\
&=&   B_{\rm pp} \bigl(-{\gammaCt}^2+\gammaCt\gammaLch + 2 \gammaCt\gammaLsp -
\gammaCs\gammaLsp \bigr) , \nonumber \\ 
\dot\gamma^{\rm C}_{\rm s} &=& - B_{\rm pp} {\gammaCs}^2 
+   {\dot\Phi}^{\rm C}_{\rm s}  \nonumber \\
&=&  B_{\rm pp} \bigl( -{\gammaCs}^2 + \gammaCs\gammaLch - 3\gammaCt\gammaLsp
\bigr) ,
\nonumber \\ 
\dot\gamma^{\rm P}_{\rm ch} &=& 2 B_{\rm ph} {\gammaPch}^2  
+ {\dot\Phi}^{\rm P}_{\rm ch} \nonumber \\ 
&=& - B_{\rm pp} \bigl( 2{\gammaPch}^2 + \gammaPch\gammaLch +
3\gammaPsp\gammaLsp \bigr)
, \nonumber \\ 
\dot\gamma^{\rm P}_{\rm sp} &=& 2 B_{\rm ph} {\gammaPsp}^2  
+ {\dot\Phi}^{\rm P}_{\rm sp} \nonumber \\ 
&=& -  B_{\rm pp} \bigl( {\gammaPsp}^2 + \gammaPch\gammaLsp +
\gammaPsp\gammaLch - 2 \gammaPsp\gammaLsp \bigr) , \nonumber \\ 
\dot\gamma^{\rm L}_{\rm ch} &=& {\dot\Phi}^{\rm L}_{\rm ch} \nonumber \\ 
&=& \half B_{\rm pp} \biggl(  \frac{3}{4}
{\gammaCt}^2 + \quarter {\gammaCs}^2 - {\gammaPch}^2 - 3 {\gammaPsp}^2 \biggr)
, \nonumber \\ 
\dot\gamma^{\rm L}_{\rm sp} &=& {\dot\Phi}^{\rm L}_{\rm sp} \nonumber \\ 
&=& \half B_{\rm pp} \biggl( \half
{\gammaCt}^2 - \half \gammaCt\gammaCs \nonumber \\ && 
- 2\gammaPch\gammaPsp + 2 {\gammaPsp}^2
+ 4 {\gammaLsp}^2 \biggr) ,
\label{gamma4_rg1D}
\eleq 
where we have used $B_{\rm pp}=-B_{\rm ph}$. $\gamma^{\rm C}$, $\gamma^{\rm
P}$ and $\gamma^{\rm L}$ are defined similarly to $\Phi^{\rm C}$, $\Phi^{\rm
P}$ and $\Phi^{\rm L}$ [Eqs.~(\ref{phiCPL1})]. Note that
$\gamma^{(4)}\dot\Pi\gamma^{(4)}$ vanishes in the Landau channel as it does
not produce any logarithmic term. Eqs.~(\ref{gamma4_rg1D}) can be simplified
by using the crossing symmetries
\bleq
\gamma^{(4)rr\bar r\bar r}_{{\rm ph},
  \sig_1\sig_2\sig_3\sig_4} &=& - 
\gamma^{(4)r\bar rr\bar r}_{{\rm
    pp},\sig_1\sig_3\sig_2\sig_4} =
\gamma^{(4)r\bar r\bar rr}_{{\rm
    pp},\sig_1\sig_3\sig_4\sig_2}  \nonumber \\ 
&=& - \gamma^{(4)r\bar r\bar rr}_{{\rm ph},
  \sig_1\sig_4\sig_3\sig_2} ,
\eleq
which lead to
\bleq
\gammaLch &=& \frac{3}{4} \gammaCt + \quarter \gammaCs 
= - \half \gammaPch - \frac{3}{2} \gammaPsp ,  \nonumber \\ 
\gammaLsp &=& \quarter \gammaCt - \quarter \gammaCs 
= - \half \gammaPch + \half \gammaPsp .
\label{gamma4_cross}
\eleq
From Eqs.~(\ref{gamma4_rg1D},\ref{gamma4_cross}), we finally deduce
\bleq
{\dot\gamma}^{\rm C}_{\rm t} &=& \frac{1}{4} B_{\rm pp}\bigl(\gammaCt-\gammaCs
\bigr)^2 , 
\nonumber \\ 
{\dot\gamma}^{\rm C}_{\rm s} &=& -\frac{3}{4}B_{\rm pp}
\bigl(\gammaCt-\gammaCs \bigr)^2,
\nonumber \\ 
{\dot\gamma}^{\rm P}_{\rm ch} &=& -\frac{3}{2} B_{\rm pp}
\bigl(\gammaPch-\gammaPsp \bigr)^2 , 
\nonumber \\ 
{\dot\gamma}^{\rm P}_{\rm sp} &=& \half B_{\rm pp} \bigl(\gammaPch-\gammaPsp
\bigr)^2, 
\nonumber \\ 
{\dot\gamma}^{\rm L}_{\rm ch} &=& 0 , 
\nonumber \\ 
{\dot\gamma}^{\rm L}_{\rm sp} &=& 4 B_{\rm pp} {\gammaLsp}^2 ,
\label{flow1D_2}
\eleq 
with the initial value $\gamma^{(4)}|_{\Lambda_0}=\Phid|_{\Lambda_0}$ given by
Eqs.~(\ref{init1D}). Alternatively, using the parameterization (\ref{init1D}),
one can rewrite Eqs.~(\ref{flow1D_2}) as two flow equations for $g_1$ and
$g_2$,  
\bleq
\dot g_1 &=& - 2 B_{\rm pp} g^2_1 , \nonumber \\
2 \dot g_2 - \dot g_1 &=& 0 .
\label{flow1D_3}
\eleq
Eqs.~(\ref{flow1D_3}) are the usual one-loop RG equations for a 1D
system.\cite{Bourbonnais95}


\begin{thebibliography}{64}
\bibitem[{not({\natexlab{c}})}]{note5}
\bibinfo{note}{$\Phi$-derivable approximations satisfy conservation laws and
  Ward identities in fermion systems.\cite{Baym62} More generally, expectation
  values of the Noether currents corresponding to global symmetries are
  conserved, but Ward identities for the self-energy and higher-order vertex
  functions can be violated, e.g. in bosonic systems with broken gauge
  invariance (see Ref.~\onlinecite{VanHees02c}).}
\bibitem[{not({\natexlab{e}})}]{note1}
\bibinfo{note}{Note that due to the antisymmetry of $G$ and $J$, the chain rule
  for derivation includes additional $\half$ factors, e.g. $\frac{\delta
  W[J]}{\delta G_\gam} = \half \sum_{\gam'} \frac{\delta J_{\gam'}}{\delta
  G_\gam} \frac{\delta W[J]}{\delta J_{\gam'}}$. For the same reason, we have
  $\frac{\delta G_{\gam_1}}{\delta G_{\gam_2}}= \frac{\delta J_{\gam_1}}{\delta
  J_{\gam_2}}=I_{\gam_1\gam_2}$ where $I$ is the unit matrix defined in
  Eq.~(\ref{Gamma2_def}).}
\bibitem[{not({\natexlab{h}})}]{note8}
\bibinfo{note}{Eq.~(\ref{gapeq111}) follows from (\ref{dotdelta}) and
  (\ref{GF_def}). To avoid ambiguities due to factors $\Theta(0)$, we use
  $\partial_\Lambda = \partial_\Lambda|_\Delta + \dot\Delta\partial_\Delta$,
  and carry out the sums over momentum before applying
  $\partial_\Lambda|_\Delta$. The same trick is used to calculate the ground
  state condensation energy in Sec.~\ref{subsubsec_gsce}.}
\bibitem[{not({\natexlab{i}})}]{note9}
\bibinfo{note}{A pairing instability in a high angular momentum channel can
  occur when the normal phase remains stable down to extremely low temperatures
  (Kohn-Luttinger effect). Such an instability would however not be
  experimentally observable.}
\end{thebibliography}

\end{document}